%% file: main.tex
\documentclass[journal]{new-aiaa} 

\usepackage[utf8]{inputenc}
\usepackage{graphicx}
\usepackage{amsfonts} 
\usepackage[version=4]{mhchem}
\usepackage{siunitx}
\usepackage{multicol}
\usepackage{longtable,tabularx}
\usepackage{diagbox}
\usepackage{booktabs}
\usepackage{comment}
\usepackage{marvosym}
\usepackage{gensymb}
\usepackage{mathbbol}
\usepackage{comment}
\usepackage{graphicx}
\usepackage{grffile}   
\usepackage{epstopdf}  
\usepackage{caption}
\usepackage{subcaption} 
\graphicspath{{Figures/}} 

\usepackage{amsthm}
\usepackage{algorithm}
\usepackage{algpseudocode}
\usepackage{float}

\input{command}

\title{Optimal Guidance with Terminal Intercept-Angle \\ 
Constraints and Acceleration Bounds} 
\author{Yoav Azaryahu\footnote{Graduate Student, The Stephen B. Klein Faculty of Aerospace Engineering, Technion, \href{mailto:yoavazaryahu@campus.technion.ac.il}{yoavazaryahu@campus.technion.ac.il}.}, Vitaly Shaferman\footnote{Associate Professor, The Stephen B. Klein Faculty of Aerospace Engineering, Technion, \href{mailto:vitalysh@technion.ac.il}{vitalysh@technion.ac.il} (Corresponding Author).}}
\affil{Technion, Israel Institute of Technology, Haifa, Israel, 3200003}

\begin{document}
\maketitle

\begin{abstract}
Terminal intercept-angle control against a maneuvering target can substantially increase the required missile acceleration, potentially leading to saturation and interception failure unless acceleration limits are explicitly addressed. The engagement is therefore formulated as a linear-quadratic optimal-control problem with bounded acceleration commands. Polynomial approximations of the line-of-sight projection coefficients are used to better represent the nonlinear engagement geometry and estimate the time-to-go.
The bounded optimal command is derived over saturated and unsaturated arcs, whose switching times are computed at each guidance step. The guidance law is derived for arbitrary linear missile dynamics and implemented for zero-order missile dynamics. For the zero-order model, the conditions under which the terminal demands can be met are derived in closed form, yielding the minimum and maximum reachable commanded terminal intercept angles. Performance is evaluated in nonlinear simulations.
Compared with its unconstrained counterparts, the bounded formulation yields substantially smaller miss distances and terminal-angle errors when saturation is encountered. Unlike corresponding bounded miss-only guidance laws, the proposed law does not reduce to its unconstrained counterpart for minimum-phase missile dynamics because the acceleration command can saturate near the end of challenging engagements. The bounded law anticipates this saturation and compensates through earlier maneuvers.

\end{abstract}

\footnotetext[0]{Presented as Paper 2027-XXXX at the AIAA SciTech Forum, Orlando, FL, 11 - 15 January 2027 (Accepted).}

\section{Introduction}
\label{Introduction}
\lettrine{M}{issiles} are often required not only to achieve a small terminal miss distance, but also to satisfy additional terminal constraints.
In particular, prescribing the terminal intercept angle can improve warhead lethality, enhance target-state observability, and allow interception from less-defended directions. However, enforcing a terminal intercept-angle constraint against a maneuvering target substantially increases the required missile acceleration. Because the interceptor has finite lateral-acceleration capability, this may lead to command saturation. In such cases, the realized trajectory deviates from the nominal trajectory predicted by the guidance law, potentially causing substantial degradation in both the miss distance and the terminal intercept-angle accuracy.

Classical interception guidance laws, including Proportional Navigation (PN) \cite{yuan_homing_1948} and Augmented Proportional Navigation (APN) \cite{garber_optimum_1968}, are mainly designed to minimize the terminal miss distance. These guidance laws impose neither a desired terminal intercept angle nor explicit missile acceleration limits. Consequently, in engagements where the terminal intercept geometry is important and the available maneuver authority is limited, alternative guidance formulations that explicitly account for these constraints are required.

Given the practical importance of terminal intercept-angle control, numerous guidance approaches have been developed over the years. These approaches generally include modified proportional-navigation guidance \cite{ratnoo2008impact,lee2013interception}, polynomial guidance \cite{lee2013polynomial}, geometric guidance \cite{tsalik2015inscribed}, nonlinear control approaches such as sliding-mode control (SMC) \cite{shima2011intercept}, optimal guidance \cite{kim_terminal_1973}, and differential-game-based guidance \cite{shaferman_linear_2008}.
Optimal guidance laws and differential-game guidance laws are particularly attractive because they are derived through optimization frameworks and are designed to minimize a specified performance index. The main distinction between the two approaches lies in how they treat the target maneuver. Optimal guidance formulations assume that the target behavior is known or predictable, resulting in a one-sided optimization problem. In contrast, differential-game formulations treat the engagement as an adversarial interaction in which both players attempt to optimize conflicting objectives. Consequently, differential-game-based guidance laws are better suited to accounting for uncertain, evading, or intelligent targets, whereas optimal guidance laws are more suitable when the target motion can be reasonably predicted.

Within the optimal-control framework, terminal-angle-constrained guidance laws have been developed over several decades. Kim and Grider \cite{kim_terminal_1973} incorporated a terminal attitude constraint into a Riccati-based optimal-control formulation. Idan, Golan, and Guelman \cite{idan_optimal_1995} obtained analytical interception laws with nonlinear kinematics and constraints on both the initial and terminal flight-path angles for targets with known trajectories. Ryoo, Cho, and Tahk \cite{ryoo_optimal_2005} extended impact-angle-constrained optimal guidance to arbitrary missile dynamics by expressing the command as a linear combination of step and ramp acceleration responses, while also proposing practical time-to-go estimation methods. In a subsequent extension, the same authors \cite{chang-kyung_ryoo_time--go_2006} introduced time-to-go power weighting in the cost function, enabling trajectory shaping through a single tuning parameter. Shaferman and Shima \cite{shaferman_linear_2008} developed closed-form optimal-control and differential-game guidance laws within a linear-quadratic framework for arbitrary-order linear missile dynamics against maneuvering targets with explicit terminal intercept-angle constraints, introducing the zero-effort angle alongside the classical zero-effort miss distance. Mishley and Shaferman \cite{mishley_linear_2022} later extended this framework to adversaries with time-varying speed profiles.

While the previous works focused primarily on terminal-angle requirements, finite interceptor maneuver capability has also been studied within the optimal-control framework. Two main approaches have been used to address acceleration constraints: hard-bound and soft-constraint formulations.

Soft formulations for addressing constant acceleration-command limits were investigated by Weiss and Shima \cite{weiss_optimal_2015}, who modified the linear-quadratic cost function by penalizing command variability without imposing an explicit hard bound. Intuitively, this additional penalty favors smoother control profiles, reducing the peak command magnitude and thereby mitigating saturation. Ben-Asher et al. \cite{ben-asher_new_2003} introduced a time-varying exponential weighting function to reflect variations in the available maneuver capability during ground-to-air engagements, where the air density decreases with altitude. Taub and Shima \cite{taub_intercept_2013} incorporated both time-varying acceleration limitations and terminal intercept-angle requirements; however, the acceleration constraint was introduced indirectly through the running cost rather than as an explicit hard bound, and iterative tuning was therefore required to achieve small miss distances and terminal-angle errors.

Explicit constant hard acceleration bounds were considered by Rusnak and Meir \cite{rusnak_optimal_1990} for engagements against maneuvering targets. They showed that, for minimum-phase missiles, the optimal law reduces to the unconstrained guidance law, whereas the nonminimum-phase case requires a more involved controller. This framework was later extended to arbitrary-order missile dynamics in \cite{rusnak_optimal_1991}. Hard time-varying acceleration bounds were later incorporated by Nahum and Shaferman \cite{nahum_optimal_2024}, who derived bounded optimal guidance laws that explicitly account for these limits for ground-to-air engagements, but without terminal intercept-angle constraints.

The preceding optimal-control-based works do not simultaneously impose a terminal intercept-angle requirement and an explicit hard acceleration bound: guidance laws that enforce a prescribed terminal intercept angle generally assume unbounded control authority, whereas formulations that account for finite acceleration either do not impose a terminal-angle constraint or incorporate the acceleration limitation indirectly through the cost function rather than as an explicit hard bound. To the best of our knowledge, no previous formulation has combined a prescribed terminal intercept angle with an explicit constant hard acceleration bound within a linear-quadratic optimal-control framework. This distinction is particularly important because enforcing terminal intercept-angle constraints is known to substantially increase the required missile acceleration \cite{shaferman_linear_2008}.

This work fills this gap by deriving a linear-quadratic optimal guidance law for interception with a prescribed terminal intercept angle under an explicit constant hard acceleration limit. To improve the prediction of the engagement evolution and the corresponding time-to-go, the line-of-sight (LOS) projection coefficients, typically assumed constant, are approximated by time-varying polynomials. The guidance law is derived for arbitrary linear missile dynamics and implemented and theoretically analyzed for zero-order missile dynamics using a cubic approximation of the projection coefficients. For the zero-order linear model, the minimum and maximum achievable intercept angles are derived in closed form. 

Unlike the classical miss-only bounded-guidance formulations of Rusnak and Meir \cite{rusnak_optimal_1990,rusnak_optimal_1991}, the introduction of a terminal intercept-angle constraint fundamentally alters the optimal guidance structure. In particular, in the miss-only case, the acceleration command magnitude for a minimum-phase missile decreases monotonically as the time-to-go decreases and converges to zero at intercept; therefore, when the target can be intercepted, saturation typically does not occur at the end of the engagement. However, when a terminal intercept-angle constraint is imposed, the optimal acceleration command generally does not converge to zero, which can lead to saturation near the end of the engagement. Consequently, the bounded controller can anticipate the upcoming saturation and compensate through earlier maneuvers during the preceding unsaturated arc. The resulting guidance law is evaluated in nonlinear simulations and compared with both the corresponding unbounded terminal-angle-constrained guidance law and the guidance law proposed in \cite{shaferman_linear_2008}. The comparison shows that the bounded guidance law yields substantially smaller miss distances and terminal-angle errors when saturation is encountered, even in scenarios involving extended periods of saturation during the engagement. An earlier version of the proposed approach was presented in \cite{Yoav2027SciTech} and is extended here.

The remainder of the paper is organized as follows. Section~\ref{sec:model_derivation}
presents the engagement model, Sec.~\ref{sec:guidance_terminal_projection} introduces the
order reduction and problem formulation, and Sec.~\ref{sec:optimal_guidance_flags}
derives the bounded optimal guidance law. Section~\ref{sec:influence_and_integrals}
specializes the law to zero-order missile dynamics and establishes the conditions under
which the terminal demands are reachable. Implementation details and simulation results are
presented in Secs.~\ref{sec:algorithm_implementation} and \ref{sec:simulation_results},
respectively. Conclusions are given in Sec.~\ref{sec:conclusions}, and additional derivations
are provided in the Appendix.

\section{Model Derivation}
\label{sec:model_derivation}

\subsection{Engagement Geometry and Nonlinear Relative Kinematics}
\label{subsec:geometry_kinematics}

\begin{figure}[hbtp]
\centering
\includegraphics[width=0.65\textwidth]{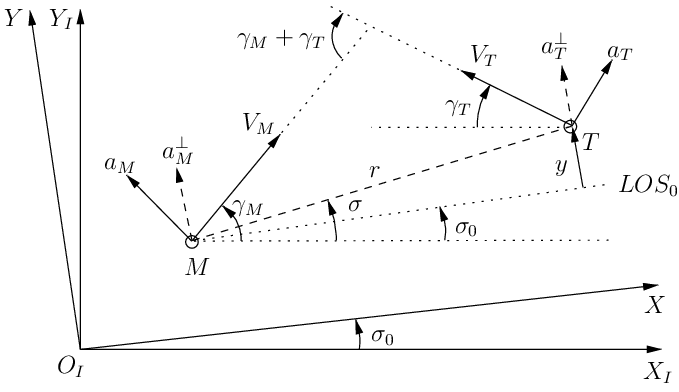}

\caption{\label{fig:Research_Proj11} Engagement geometry and relative motion variables.}
\end{figure}

We consider a planar missile-target engagement, illustrated in Fig.~\ref{fig:Research_Proj11}. 
The missile and target are denoted by the subscripts \(M\) and \(T\), respectively. 
Their speeds, lateral accelerations, and flight-path angles are denoted by \(V\), \(a\), and \(\gamma\), respectively. 
The relative displacement is described in polar coordinates \((r,\sigma)\), where \(r\) is the relative range and \(\sigma\) is the line-of-sight (LOS) angle measured from the inertial \(X_I\)-axis. 
Both vehicles are assumed to maneuver by applying lateral acceleration perpendicular to their respective velocity vectors.

From the nonlinear relative kinematics in polar coordinates, the range-rate along the LOS is
\begin{equation}
\label{eq:rdot}
\dot{r}
=
-\Bigl[
V_M \cos(\gamma_M-\sigma)
+
V_T \cos(\gamma_T+\sigma)
\Bigr]
\end{equation}
and the relative transverse velocity is
\begin{equation}
\label{eq:sigmadot}
r\dot{\sigma}
=
- V_M \sin(\gamma_M-\sigma)
+
V_T \sin(\gamma_T+\sigma)
\end{equation}

The missile and target are assumed to move at constant speeds \(V_M\) and \(V_T\), so that their flight-path angle rates satisfy
\begin{equation}
\label{eq:gammadot}
\dot{\gamma}_M=\frac{a_M}{V_M},
\quad
\dot{\gamma}_T=\frac{a_T}{V_T}
\end{equation}

\subsection{Linearization Model for Guidance Law Derivation}
\label{subsec:linearization_collision_geometry}

To obtain a tractable guidance model, the nonlinear engagement kinematics are linearized about a nominal collision geometry. 
Let \(LOS_0\) denote the initial line of sight, and let \(y\) be the lateral displacement measured perpendicular to \(LOS_0\). 
Under the standard small-deviation assumption, namely that both vehicles remain sufficiently close to the reference collision triangle during the engagement, the lateral relative dynamics may be approximated by
\begin{equation}
\label{eq:y_ddot}
\ddot{y}=a_T^\perp-a_M^\perp
\end{equation}
where \(a_T^\perp\) and \(a_M^\perp\) are the target and missile acceleration components projected onto the direction perpendicular to \(LOS_0\).

The time-to-go is defined by
\begin{equation}
\label{eq:tgo_def}
t_{go}\triangleq t_f-t
\end{equation}
where \(t_f\) denotes the final interception time. 
The perpendicular acceleration components are modeled as
\begin{equation}
\label{eq:perp_accels}
a_T^\perp = K_T(t_{go})\,a_T,
\quad
a_M^\perp = K_M(t_{go})\,a_M
\end{equation}
where \(K_T(t_{go})\) and \(K_M(t_{go})\) are the time-varying geometric projection terms assumed to be known. The present formulation extends the classical constant \(K_T\) and \(K_M\) projection terms to provide a more accurate representation of the nonlinear engagement geometry and the corresponding time-to-go approximation. 

To preserve a fully analytic derivation, these geometry terms are approximated by polynomials in \(t_{go}\)
\begin{equation}
\label{eq:KM_KT_polynomials}
K_M(t_{go})=\sum_{i=0}^{n_M} k_{M,i}\,t_{go}^i,
\quad
K_T(t_{go})=\sum_{i=0}^{n_T} k_{T,i}\,t_{go}^i
\end{equation}
This representation allows the influence functions and the associated interval integrals to be evaluated in closed form. 
The constant projection case is recovered as a special case by choosing \(n_M=n_T=0\). The proposed approach for estimating these terms is presented later in the paper.

The missile and target dynamics are represented by Linear Time-Invariant (LTI), arbitrary-order models. Their state-space representations are written as
\begin{subequations}
\label{eq:internal_dynamics}
\begin{align} \label{eq:missile_dynamics}
\dot{\bfx}_M &= \bfA_M \bfx_M + \bfb_M u,
\quad
a_M = \bfc_M^T \bfx_M + d_M u \\  \label{eq:target_dynamics}
\dot{\bfx}_T &= \bfA_T \bfx_T + \bfb_T w,
\quad
a_T = \bfc_T^T \bfx_T + d_T w
\end{align}
\end{subequations}
where \(\bfx_M\) and \(\bfx_T\) are the missile and target internal-state vectors, respectively, and \(u\) and \(w\) are the missile and target acceleration commands, respectively.

The overall state vector is defined as
\begin{equation}
\label{eq:state_vector}
\bfx
\triangleq
\begin{bmatrix}
y & \dot{y} & \Delta\gamma & \bfx_T^T & \bfx_M^T
\end{bmatrix}^T, \quad \Delta\gamma \triangleq \gamma_T+\gamma_M
\end{equation}
where the relative intercept angle \(\Delta\gamma\) is included explicitly because it is
used to impose the terminal intercept-angle requirement.

Using \eqref{eq:y_ddot}, \eqref{eq:perp_accels}, and \eqref{eq:internal_dynamics}, the Equations Of Motion (EOM) can be written in the matrix  Linear Time-Varying (LTV) form
\begin{equation}
\label{eq:LTV_model}
\dot{\bfx}=\bfA(t_{go})\,\bfx+\bfb(t_{go})\,u+\bfc(t_{go})\,w
\end{equation}
where
\begin{equation}
\label{eq:LTV_matrices}
\bfA(t_{go})=
\begin{bmatrix}
0 & 1 & 0 & \mathbf{0} & \mathbf{0} \\
0 & 0 & 0 & K_T(t_{go})\,\bfc_T^T & -K_M(t_{go})\,\bfc_M^T \\
0 & 0 & 0 & \frac{1}{V_T}\bfc_T^T & \frac{1}{V_M}\bfc_M^T \\
\mathbf{0} & \mathbf{0} & \mathbf{0} & \bfA_T & \mathbf{0} \\
\mathbf{0} & \mathbf{0} & \mathbf{0} & \mathbf{0} & \bfA_M
\end{bmatrix}, \quad 
\bfb(t_{go})=
\begin{bmatrix}
0 \\
-K_M(t_{go})\,d_M \\
\frac{d_M}{V_M} \\
\mathbf{0} \\
\bfb_M
\end{bmatrix}, \quad
\bfc(t_{go})=
\begin{bmatrix}
0 \\
K_T(t_{go})\,d_T \\
\frac{d_T}{V_T} \\
\bfb_T \\
\mathbf{0}
\end{bmatrix}
\end{equation}
and \(\mathbf{0}\) denotes a zero block of appropriate dimension.

\section{Guidance Problem Formulation and Order Reduction}
\label{sec:guidance_terminal_projection}

\subsection{Optimization Problem}
\label{subsec:cost_terminal_angle}

To achieve interception with a prescribed terminal intercept angle, we consider the quadratic cost
\begin{equation}
\label{eq:J_tv}
\mathcal{J}
=
\frac{1}{2}a\,x_1^2(t_f)
+
\frac{1}{2}b\bigl[x_3(t_f)-\chi_c\bigr]^2
+
\frac{1}{2}\int_t^{t_f} u^2(\xi)\,d\xi
\end{equation}
with non-negative weights $a,b$, required terminal intercept angle $\chi_c$, and the bounded control
\begin{equation}
\label{eq:u_bound}
|u|\leq u_{\max}
\end{equation}
In the limit $a\to\infty$, perfect interception is enforced, whereas $b\to\infty$ enforces the prescribed intercept angle.

\subsection{Coordinate Change and Order Reduction}
\label{subsec:tv_terminal_projection_main}
The order of the problem is reduced through the ``Terminal-Projection'' transformation based on the solution of the differential equation in \eqref{eq:LTV_model}. 
Let $\bfPhi(t_f,t)$ be the state transition matrix associated with \eqref{eq:LTV_model}, satisfying
\begin{equation}
\label{eq:stm_def_main}
\dot{\bfPhi}(t_f,t)=-\bfPhi(t_f,t)\,\bfA(t),
\quad
\bfPhi(t,t)=\bfI
\end{equation}
Assuming that the target's future maneuver strategy is known, the projected variables are defined as
\begin{equation}
\label{eq:Z_def_main}
\bfZ(t)=\bfE\,\bfPhi(t_f,t)\,\bfx(t)+
\bfE\int_t^{t_f}\bfPhi(t_f,\tau)\,\bfc(\tau)\,w(\tau)\,d\tau, \quad\bfE=
\begin{bmatrix}
1 & 0 & 0 & \mathbf{0} & \mathbf{0} \\
0 & 0 & 1 & \mathbf{0} & \mathbf{0}
\end{bmatrix}
\end{equation}
where the selection matrix $\bfE$ extracts $x_1$ and $x_3$. Since $\bfPhi(t_f,t_f)=\bfI$, substituting $t=t_f$ yields
\begin{equation}
\label{eq:Z_terminal_property}
Z_1(t_f)=x_1(t_f),\quad Z_2(t_f)=x_3(t_f)
\end{equation}
so $Z_1$ is the zero-effort-miss (ZEM) and $Z_2$ is the zero-effort-angle (ZEA), i.e., the values of $x_1(t_f)$ and $x_3(t_f)$ obtained under $u\equiv 0$ and the assumed target maneuver. Substituting \eqref{eq:Z_terminal_property}, the cost function in \eqref{eq:J_tv} becomes
\begin{equation}
\label{eq:J_in_Z_tv}
\mathcal{J}
=
\frac{1}{2}a\,Z_1^2(t_f)
+
\frac{1}{2}b\bigl[Z_2(t_f)-\chi_c\bigr]^2
+
\frac{1}{2}\int_t^{t_f} u^2(\xi)\,d\xi
\end{equation}

Differentiating \eqref{eq:Z_def_main} and using \eqref{eq:stm_def_main} together with the state equation \eqref{eq:LTV_model}, yields
\begin{equation}
\label{eq:Zdot_reduced_main}
\dot{\bfZ}=\tilde{\bfB}(t_{go})\,u , \quad
\tilde{\bfB}(t_{go})=\begin{bmatrix}\tilde B_1(t_{go}) \\ \tilde B_2(t_{go})\end{bmatrix}\triangleq\bfE\,\bfPhi(t_f,t)\,\bfb(t)
\end{equation}

The resulting reduced-order dynamics are independent of the state and target acceleration command, with dependence only on $t_{go}$ and the control input $u$.
With the polynomial approximations of $K_M(t_{go})$ and $K_T(t_{go})$ in \eqref{eq:KM_KT_polynomials}, the influence functions are
\begin{equation}\label{eq:B1B2_closed_form_main}
\tilde B_1(t_{go})=
-\sum_{i=0}^{n_M}k_{M,i}(i+1)!\,\mathcal{L}^{-1}_{t_{go}}\!\left[\frac{G_M(s)}{s^{i+2}}\right], \quad
\tilde B_2(t_{go})=
\frac{1}{V_M}\,\mathcal{L}^{-1}_{t_{go}}\!\left[\frac{G_M(s)}{s}\right]
\end{equation}
where $\mathcal{L}^{-1}_{t_{go}}$ is the inverse Laplace operator to $t_{go}$, and $G_M(s)$ is the missile acceleration transfer function given by
\begin{equation}
\label{eq:GM_def}
G_M(s)\triangleq\bfc_M^T(s\bfI-\bfA_M)^{-1}\bfb_M+d_M
\end{equation}
For brevity and to improve readability, the derivation of \eqsref{eq:B1B2_closed_form_main}
is given in \ref{app:reduced_order_ZEV}.

\section{Bounded Optimal Guidance Law}
\label{sec:optimal_guidance_flags}

\subsection{Unconstrained Optimal Controller}
\label{subsec:unconstrained_command}

Using \eqref{eq:J_in_Z_tv} and \eqref{eq:Zdot_reduced_main}, the Hamiltonian is
\begin{equation}
\label{eq:H_general}
\mathcal{H}=\frac{1}{2}u^2+\lambda_1\tilde B_1(t_{go})\,u+\lambda_2\tilde B_2(t_{go})\,u
\end{equation}
Because $\mathcal{H}$ does not explicitly depend on the states $Z_1$ and $Z_2$, the costates are constant
\begin{equation}
\label{eq:costates_constant}
\dot\lambda_i=-\frac{\partial\mathcal{H}}{\partial Z_i}=0,\quad i\in\{1,2\}
\end{equation}
and the terminal conditions fix their values
\begin{equation}
\label{eq:costate_terminal}
\lambda_1(t)=\lambda_1(t_f)=a\,Z_1(t_f),\quad
\lambda_2(t)=\lambda_2(t_f)=b\bigl[Z_2(t_f)-\chi_c\bigr]
\end{equation}
Substituting \eqref{eq:costate_terminal} into \eqref{eq:H_general} the optimal unbounded controller satisfies 
\begin{equation}
\label{eq:u0_general}
\frac{\partial\mathcal{H}}{\partial u}=0 \quad
\Rightarrow \quad 
u(\xi)=-a\,Z_1(t_f)\,\tilde B_1(\xi)-b\bigl[Z_2(t_f)-\chi_c\bigr]\tilde B_2(\xi), \quad \xi\in[0,t_{go}]
\end{equation}

Integrating the reduced-order dynamics \eqref{eq:Zdot_reduced_main} over the engagement horizon yields
\begin{equation}
\label{eq:Z_tf_integral_unsat}
Z_i(t_f)=Z_i(t)+\int_0^{t_{go}}\tilde B_i(\xi)\,u(\xi)\,d\xi,\quad i\in\{1,2\}
\end{equation}
Substituting \eqref{eq:u0_general}, rearranging, and using the integral definitions
\begin{equation}
\label{eq:I_defs}
\mathcal{I}_{ij}(t_{go})\triangleq\int_0^{t_{go}}\tilde B_i(\xi)\,\tilde B_j(\xi)\,d\xi,\quad i,j\in\{1,2\}
\end{equation}
yields the system
\begin{subequations}
\label{eq:Z_system_unsat}
\begin{equation}
\bigl[1+a\mathcal{I}_{11}(t_{go})\bigr]Z_1(t_f)+b\,\mathcal{I}_{12}(t_{go})\bigl[Z_2(t_f)-\chi_c\bigr]=Z_1(t)
\label{eq:Z1_system_unsat}
\end{equation}
\begin{equation}
a\,\mathcal{I}_{12}(t_{go})\,Z_1(t_f)+\bigl[1+b\mathcal{I}_{22}(t_{go})\bigr]\bigl[Z_2(t_f)-\chi_c\bigr]=Z_2(t)-\chi_c
\label{eq:Z2_system_unsat}
\end{equation}
\end{subequations}
whose solution is
\begin{subequations}
\label{eq:Z_tf_closed_unsat}
\begin{equation}
Z_1(t_f)=\frac{\bigl[1+b\mathcal{I}_{22}(t_{go})\bigr]Z_1(t)-b\,\mathcal{I}_{12}(t_{go})\bigl[Z_2(t)-\chi_c\bigr]}{\Delta_z}
\label{eq:Z1_tf_closed_unsat}
\end{equation}
\begin{equation}
Z_2(t_f)-\chi_c=\frac{\bigl[1+a\mathcal{I}_{11}(t_{go})\bigr]\bigl[Z_2(t)-\chi_c\bigr]-a\,\mathcal{I}_{12}(t_{go})\,Z_1(t)}{\Delta_z}
\label{eq:Z2_tf_closed_unsat}
\end{equation}
\end{subequations}
where
\begin{equation}
\label{eq:Delta_2}
\Delta_z\triangleq\bigl[1+a\mathcal{I}_{11}(t_{go})\bigr]\bigl[1+b\mathcal{I}_{22}(t_{go})\bigr]-ab\,\mathcal{I}_{12}^2(t_{go})
\end{equation}
 and by the Cauchy-Schwarz inequality ${\mathcal{I}}_{11}{\mathcal{I}}_{22} \ge {\mathcal{I}}_{12}^{\,2}$, therefore, ${\Delta}_z>0 \; \forall t_{go}$. 
Implementing the controller also requires the ZEM and ZEA projected variables. For a constant target acceleration these are
\begin{subequations}
\label{eq:Z_proj_constaT}
\begin{equation}
Z_1=x_1+t_{go}\,x_2-\sum_{i=0}^{n_M}k_{M,i}(i+1)!\,\mathcal{L}^{-1}_{t_{go}}\!\left[\frac{\bfc_M^T(s\bfI-\bfA_M)^{-1}}{s^{i+2}}\right]\bfx_M+\sum_{i=0}^{n_T}\frac{k_{T,i}\,a_T}{i+2}\,t_{go}^{\,i+2}
\label{eq:Z1_constaT}
\end{equation}
\begin{equation}
Z_2=x_3+\frac{1}{V_M}\,\mathcal{L}^{-1}_{t_{go}}\!\left[\frac{\bfc_M^T(s\bfI-\bfA_M)^{-1}}{s}\right]\bfx_M+\frac{a_T\,t_{go}}{V_T}
\label{eq:Z2_constaT}
\end{equation}
\end{subequations}
the derivation of which is given in \ref{app:reduced_order_ZEV}. Substituting the terminal values \eqref{eq:Z_tf_closed_unsat} and the projected variables \eqref{eq:Z_proj_constaT} into \eqref{eq:u0_general} yields the optimal unconstrained controller.

\subsection{Bounded Optimal Controller}
\label{subsec:bounded_command_switching}

The cost in \eqref{eq:J_in_Z_tv} is minimized over the admissible set \(|u|\le u_{\max}\). The bounded optimum is the pointwise saturation of the unconstrained command structure
\begin{equation}
\label{eq:bounded_command_general}
\bar u(\xi)=u_{\max}\,\mathrm{sat}\!\left[\frac{\bar u_0(\xi)}{u_{\max}}\right],
\quad
\bar u_0(\xi)\triangleq -a\,\bar Z_1(t_f)\,\tilde B_1(\xi)-b\,\bigl[\bar Z_2(t_f)-\chi_c\bigr]\tilde B_2(\xi)
\end{equation}
where \(\bar{\Box}\) indicates values in the bounded problem, \(\bar u_0\) is the unsaturated bounded controller, \(\bar Z_1(t_f)\) and \(\bar Z_2(t_f)\) are the terminal ZEM and ZEA in the bounded problem, and \(\mathrm{sat}\) is the saturation function. Note that the terminal ZEM and ZEA in the bounded problem might differ from those in the unbounded problem.

\subsubsection{Saturated and unsaturated arcs and terminal-value system}
\label{subsubsec:arcs_additivity}
The saturation structure over the interval \([0,t_{go}]\) follows from \(\bar u_0\), which is determined by which constraint, if any, is active
\begin{equation}
\label{eq:region_defs}
\mathcal{R}_S^{+}=\bigl\{\xi:\bar u_0(\xi)\ge +u_{\max}\bigr\},
\quad
\mathcal{R}_S^{-}=\bigl\{\xi:\bar u_0(\xi)\le -u_{\max}\bigr\},
\quad
\mathcal{R}_U=\bigl\{\xi:|\bar u_0(\xi)|<u_{\max}\bigr\}
\end{equation}
and cover \([0,t_{go}]\) without overlap, with the corresponding command
\begin{equation}
\label{eq:command_by_region}
\bar u(\xi)=
\begin{cases}
+u_{\max} & \xi\in\mathcal{R}_S^{+} \\
-u_{\max} & \xi\in\mathcal{R}_S^{-} \\
\bar u_0(\xi) & \xi\in\mathcal{R}_U
\end{cases}
\end{equation}

Since \(\tilde B_1\) and \(\tilde B_2\) are continuous and the terminal values \(\bar Z_1(t_f)\), \(\bar Z_2(t_f)\) are fixed, \(\bar u_0\) is continuous in \(\xi\). It can therefore pass between the unsaturated region \(\mathcal{R}_U\) and a saturated region only where it reaches the bound: it enters \(\mathcal{R}_S^{+}\) where \(\bar u_0=+u_{\max}\) and \(\mathcal{R}_S^{-}\) where \(\bar u_0=-u_{\max}\). It cannot pass directly between \(\mathcal{R}_S^{+}\) and \(\mathcal{R}_S^{-}\), since a continuous \(\bar u_0\) going from \(+u_{\max}\) to \(-u_{\max}\) must cross the unsaturated region \(\mathcal{R}_U\) in between.

Each region is therefore a finite union of disjoint sub-intervals of $[0,t_{go}]$, referred to as arcs, whose endpoints are the points where $\bar u_0$ reaches $\pm u_{\max}$. Since integration is additive over disjoint intervals, the contribution of a region to any terminal integral is the sum of the contributions of its arcs, regardless of their number or location.

A saturated arc carries a constant command, which factors out of the terminal integrals and leaves only $\tilde B_i$. Define the saturated-arc integrals
\begin{equation}
\label{eq:Q_def}
\mathcal{Q}_i(t_{go})\triangleq\int_0^{t_{go}}\tilde B_i(\xi)\,d\xi,\quad i\in\{1,2\}
\end{equation}
For any region \(\mathcal{R}\subseteq[0,t_{go}]\) made up of disjoint sub-intervals \(\{[\tau_k^{\mathrm{lo}},\tau_k^{\mathrm{up}}]\}_k\), additivity gives
\begin{equation}
\label{eq:additivity_B}
\int_{\mathcal{R}}\tilde B_i(\xi)\,d\xi=\sum_k\bigl[\mathcal{Q}_i(\tau_k^{\mathrm{up}})-\mathcal{Q}_i(\tau_k^{\mathrm{lo}})\bigr], \quad
\int_{\mathcal{R}}\tilde B_i(\xi) \tilde B_j(\xi) \,d\xi=\sum_k\bigl[\mathcal{I}_{ij}(\tau_k^{\mathrm{up}})-\mathcal{I}_{ij}(\tau_k^{\mathrm{lo}})\bigr],
\quad i,j\in\{1,2\}
\end{equation}

Integrating the reduced-order dynamics over \([0,t_{go}]\) and grouping the result by region yields
\begin{equation}
\label{eq:Z_region_grouped}
\bar Z_i(t_f)=Z_i(t)
+u_{\max}\!\int_{\mathcal{R}_S^{+}}\!\tilde B_i\,d\xi
-u_{\max}\!\int_{\mathcal{R}_S^{-}}\!\tilde B_i\,d\xi
+\int_{\mathcal{R}_U}\!\tilde B_i\,\bar u_0\,d\xi,
\quad i\in\{1,2\}
\end{equation}
On a saturated arc, the command equals the constant \(\pm u_{\max}\). The saturated contributions therefore depend only on the switching times and are evaluated through \eqref{eq:additivity_B}
\begin{equation}
\label{eq:S_def}
\mathcal{S}_i\triangleq u_{\max}\!\left[\int_{\mathcal{R}_S^{+}}\!\tilde B_i\,d\xi-\int_{\mathcal{R}_S^{-}}\!\tilde B_i\,d\xi\right],
\quad i\in\{1,2\}
\end{equation}
so that the terminal values in \eqref{eq:Z_region_grouped} become \(\bar Z_i(t_f)=Z_i(t)+\mathcal{S}_i+\int_{\mathcal{R}_U}\tilde B_i\,\bar u_0\,d\xi\).

On \(\mathcal{R}_U\) the command equals \(\bar u_0\), which is linear in the unknown terminal values. Define the product integrals over the unsaturated region
\begin{equation}
\label{eq:Ihat_defs}
\bar{\mathcal{I}}_{ij}\triangleq\int_{\mathcal{R}_U}\tilde B_i\,\tilde B_j\,d\xi,
\quad i,j\in\{1,2\}
\end{equation}
obtained from \(\mathcal{I}_{ij}\) over the sub-intervals of \(\mathcal{R}_U\) by the additivity of \eqref{eq:additivity_B}.

Substituting \(\bar u_0\) into \eqref{eq:Z_region_grouped}, using \eqref{eq:Ihat_defs}, and collecting terms in \(\bar Z_1(t_f)\) and \(\bar Z_2(t_f)\) gives the linear system
\begin{subequations}
\label{eq:Z_system_general}
\begin{equation}
\bigl[1+a\bar{\mathcal{I}}_{11}\bigr]\bar Z_1(t_f)+b\bar{\mathcal{I}}_{12}\bigl[\bar Z_2(t_f)-\chi_c\bigr]=Z_1(t)+\mathcal{S}_1
\label{eq:Z1_system_general}
\end{equation}
\begin{equation}
a\bar{\mathcal{I}}_{12}\,\bar Z_1(t_f)+\bigl[1+b\bar{\mathcal{I}}_{22}\bigr]\bigl[\bar Z_2(t_f)-\chi_c\bigr]=Z_2(t)-\chi_c+\mathcal{S}_2
\label{eq:Z2_system_general}
\end{equation}
\end{subequations}
whose solution is
\begin{subequations}
\label{eq:Z_tf_general}
\begin{equation}
\bar Z_1(t_f)=\frac{\bigl[1+b\bar{\mathcal{I}}_{22}\bigr]\bigl[Z_1(t)+\mathcal{S}_1\bigr]-b\bar{\mathcal{I}}_{12}\bigl[Z_2(t)-\chi_c+\mathcal{S}_2\bigr]}{\bar{\Delta}_z}
\label{eq:Z1_tf_general}
\end{equation}
\begin{equation}
\bar Z_2(t_f)-\chi_c=\frac{\bigl[1+a\bar{\mathcal{I}}_{11}\bigr]\bigl[Z_2(t)-\chi_c+\mathcal{S}_2\bigr]-a\bar{\mathcal{I}}_{12}\bigl[Z_1(t)+\mathcal{S}_1\bigr]}{\bar{\Delta}_z}
\label{eq:Z2_tf_general}
\end{equation}
\end{subequations}
where
\begin{equation}
\label{eq:Delta_general}
\bar{\Delta}_z \triangleq\bigl[1+a\bar{\mathcal{I}}_{11}\bigr]\bigl[1+b\bar{\mathcal{I}}_{22}\bigr]-ab\,\bar{\mathcal{I}}_{12}^{\,2}
\end{equation}
 and by the Cauchy-Schwarz inequality $\bar{\mathcal{I}}_{11}\bar{\mathcal{I}}_{22} \ge \bar{\mathcal{I}}_{12}^{\,2}$, therefore, $\bar{\Delta}_z>0$. These expressions hold for any number of saturated arcs of either sign.

\subsubsection{Switching times and arc sequence}
\label{subsubsec:switching_times}

The switching times are the points in \([0,t_{go}]\) where \(\bar u_0\) reaches \(\pm u_{\max}\), that is, \(\xi_{j}\) is the $j$-th root of
\begin{equation}
\label{eq:switch_equation_general}
a\,\bar Z_1(t_f)\,\tilde B_1(\xi_{j})+b\,\bigl[\bar Z_2(t_f)-\chi_c\bigr]\tilde B_2(\xi_{j})+S_j\,u_{\max}=0, \quad j\in\{1,...,m\},
\quad S_j\in\{+1,-1\}, \quad S=[S_1,...,S_m]
\end{equation}
with \(S_j=+1\) at a boundary of \(\mathcal{R}_S^{+}\) and \(S_j=-1\) at a boundary of \(\mathcal{R}_S^{-}\). The number of crossings, $m$, and hence the number of saturated and unsaturated arcs, is not fixed in advance: it is determined by \(\bar u_0\), which depends in turn on the regions through \(\mathcal{S}_1\), \(\mathcal{S}_2\), and \(\bar{\mathcal{I}}_{ij}\). The terminal values \eqref{eq:Z_tf_general} and the switching times \eqref{eq:switch_equation_general} are therefore coupled and are solved together; the solution procedure for any linear missile dynamics and geometry-coefficient approximation is given in Sec.~\ref{sec:algorithm_implementation}. The maximum number of crossings is set by how many times \(\bar u_0\) can intersect \(\pm u_{\max}\), which depends on the missile dynamics and the geometry-coefficient approximation. Higher-order dynamics and geometry-coefficient approximations may therefore admit more saturated arcs than the zero-order case and a lower-order geometry-coefficient approximation.

An arc sequence is an ordered sequence of unsaturated, positively saturated, and negatively saturated arcs. An arc sequence and a command are consistent when, on each arc, the unsaturated command lies on the side of the bound specified by that arc:
\begin{equation}
\label{eq:admissible}
|\bar u_0(\xi)|\le u_{\max}\ \text{on every unsaturated arc},\qquad
S_j\,\bar u_0(\xi)\ge u_{\max}\ \text{on every saturated arc } j
\end{equation}

Unfortunately, the sequence is not known in advance, so \eqref{eq:admissible} cannot be imposed while solving; it is verified afterward on each solved candidate.

\section{Zero-Order Missile Dynamics with Cubic Geometry
Coefficients}
\label{sec:influence_and_integrals}

This section specializes the bounded guidance law of
Sec.~\ref{sec:optimal_guidance_flags} to zero-order missile dynamics with cubic geometry
coefficients. This is done to simplify and analyze the guidance law. We first derive the influence functions, integrals, and zero-effort variables. The bounded command structure is then presented under simplifying assumptions, followed by the conditions under which the missile can intercept the target at the commanded terminal angle.

\subsection{Zero-Order Dynamics and Integrals}
\label{subsec:zo_dynamics_integrals}

For zero-order missile dynamics \eqref{eq:missile_dynamics} and \eqref{eq:GM_def} reduce to
\begin{equation}
\label{eq:zero_order_dyn}
a_M=u \quad \Rightarrow \quad
\bfA_M=0, \; \bfb_M=0, \; \bfc_M=0, \; d_M=1  \quad \Rightarrow \quad
G_M(s)=1
\end{equation}
Substituting into \eqref{eq:B1B2_closed_form_main} and using the inverse Laplace identity
$\mathcal{L}^{-1}_{t_{go}}\!\left[\frac{1}{s^{i+2}}\right]=\frac{t_{go}^{\,i+1}}{(i+1)!}$
the $(i+1)!$ factors cancel, yielding
\begin{equation}
\label{eq:B1B2_zero_order}
\tilde B_1(\xi)=-\xi\,K_M(\xi),
\quad
\tilde B_2(\xi)=\frac{1}{V_M}
\end{equation}
The geometry coefficients $K_M(\xi)$ and $K_T(\xi)$ are approximated by the cubic
polynomials
\begin{equation}
\label{eq:K_cubic_app}
K_M(\xi)=k_{M,0}+k_{M,1}\xi+k_{M,2}\xi^2+k_{M,3}\xi^3, \quad
K_T(\xi)=k_{T,0}+k_{T,1}\xi+k_{T,2}\xi^2+k_{T,3}\xi^3
\end{equation}
so that
\begin{equation}
\label{eq:B1B2_cubic}
\tilde B_1(\xi)=-\bigl(k_{M,0}\xi+k_{M,1}\xi^2+k_{M,2}\xi^3+k_{M,3}\xi^4\bigr),
\quad
\tilde B_2(\xi)=\frac{1}{V_M}
\end{equation}
and the zero-effort variables in \eqref{eq:Z_proj_constaT} reduce to
\begin{equation}
\label{eq:Z_zero_order}
Z_1(t)=y+t_{go}\,\dot y+\sum_{i=0}^{3}\frac{k_{T,i}\,a_T}{i+2}\,t_{go}^{\,i+2}, \quad
Z_2(t)=\Delta\gamma+\frac{a_T\,t_{go}}{V_T}
\end{equation}
Substituting \eqref{eq:B1B2_cubic} into the saturated-arc integrals  \eqref{eq:Q_def} gives

\begin{equation}
\label{eq:Q1Q2_zero}
\mathcal{Q}_1(t_{go}) = -\left[k_{M,0}\frac{t_{go}^2}{2}+
k_{M,1}\frac{t_{go}^3}{3}+k_{M,2}\frac{t_{go}^4}{4}+
k_{M,3}\frac{t_{go}^5}{5}\right], \quad
\mathcal{Q}_2(t_{go})=\frac{t_{go}}{V_M}
\end{equation}
and substituting the same expressions into \eqref{eq:I_defs} gives the product integrals
\begin{subequations}
\label{eq:Iij_zero}
\begin{align}\label{eq:I11_zero}
\mathcal{I}_{11}(t_{go})={}&k_{M,0}^2\frac{t_{go}^3}{3}+
2k_{M,0}k_{M,1}\frac{t_{go}^4}{4}+ \bigl(2k_{M,0}k_{M,2}+k_{M,1}^2\bigr)\frac{t_{go}^5}{5}
\notag\\
&+
\bigl(2k_{M,0}k_{M,3}+2k_{M,1}k_{M,2}\bigr)\frac{t_{go}^6}{6}
+
\bigl(2k_{M,1}k_{M,3}+k_{M,2}^2\bigr)\frac{t_{go}^7}{7}
+
2k_{M,2}k_{M,3}\frac{t_{go}^8}{8}
+
k_{M,3}^2\frac{t_{go}^9}{9}
\end{align}
\begin{equation} \label{eq:I12_zero}
\mathcal{I}_{12}(t_{go})=-\frac{1}{V_M}\left[k_{M,0}\frac{t_{go}^2}{2}+
k_{M,1}\frac{t_{go}^3}{3}+k_{M,2}\frac{t_{go}^4}{4}+k_{M,3}\frac{t_{go}^5}{5}\right], \quad
\mathcal{I}_{22}(t_{go})=\frac{t_{go}}{V_M^2}
\end{equation}
\end{subequations}

\subsection{Typical Bounded Command Arcs}
\label{subsec:zo_arcs}

For zero-order dynamics \(\tilde B_2=1/V_M\) is constant and
\(\tilde B_1(\xi)=-\xi K_M(\xi)\) is a fourth-order polynomial under the cubic geometry approximation in \eqref{eq:K_cubic_app}, vanishing at $\xi=0$.

By \eqref{eq:bounded_command_general}, the unsaturated command $\bar u_0$ is $\tilde B_1$ scaled by a constant and biased by another constant, hence a fourth-order polynomial in $\xi$. Each of $\bar u_0=\pm u_{\max}$ then has up to four roots in $[0,t_{go}]$, and the command could switch up to eight times.

To simplify the analysis in this section, $\tilde B_1$ is assumed strictly monotonic over $[0,t_{go}]$. A scaled strictly monotonic function plus a bias is monotonic, so $\bar u_0$ is monotonic and reaches each bound at most once, giving at most two switching times $t_{go_{s1}}$ and $t_{go_{s2}}$. The assumption holds when

\begin{equation}
\label{eq:monotonicity_condition}
\frac{d \tilde B_1(\xi)}{d\xi}= \tilde B_1'(\xi)=-\bigl(\xi K_M(\xi)\bigr)'=-\bigl[k_{M,0}+2k_{M,1}\xi+3k_{M,2}\xi^2+4k_{M,3}\xi^3\bigr]
\end{equation}

does not change sign over $[0,t_{go}]$, which can be easily verified. 
Since $\tilde B_1(0)=0$, a strictly monotonic $\tilde B_1$ has one sign over $[0,t_{go}]$, a property used in Sec.~\ref{subsec:reach_zero_order}.

It should be noted that condition \eqref{eq:monotonicity_condition} is satisfied by the constant-$K_M$ geometry approximation commonly used in the guidance literature. Under this approximation, $\tilde B_1$ is linear in $\xi$ and hence strictly monotonic. The higher-order terms in the cubic fit add smaller corrections to this dominant constant term. Moreover, condition \eqref{eq:monotonicity_condition} was satisfied in all engagements examined in Sec.~\ref{sec:simulation_results}.

If the condition in \eqref{eq:monotonicity_condition} does not hold, one can reduce the approximation order of $K_M$ up to zero, for which this condition is guaranteed. 
Importantly, the general derivation in Sec.~\ref{sec:optimal_guidance_flags}, the algorithm of Sec.~\ref{sec:algorithm_implementation}, and the simulations in Sec.~\ref{sec:simulation_results} do not rely on this condition, and it is only used for the presentation of the typical arc patterns of the controller in this section and the reachability analysis in Sec.~\ref{subsec:reach_zero_order}.

\begin{figure}[hbtp]
\centering
\includegraphics[width=0.65\textwidth]{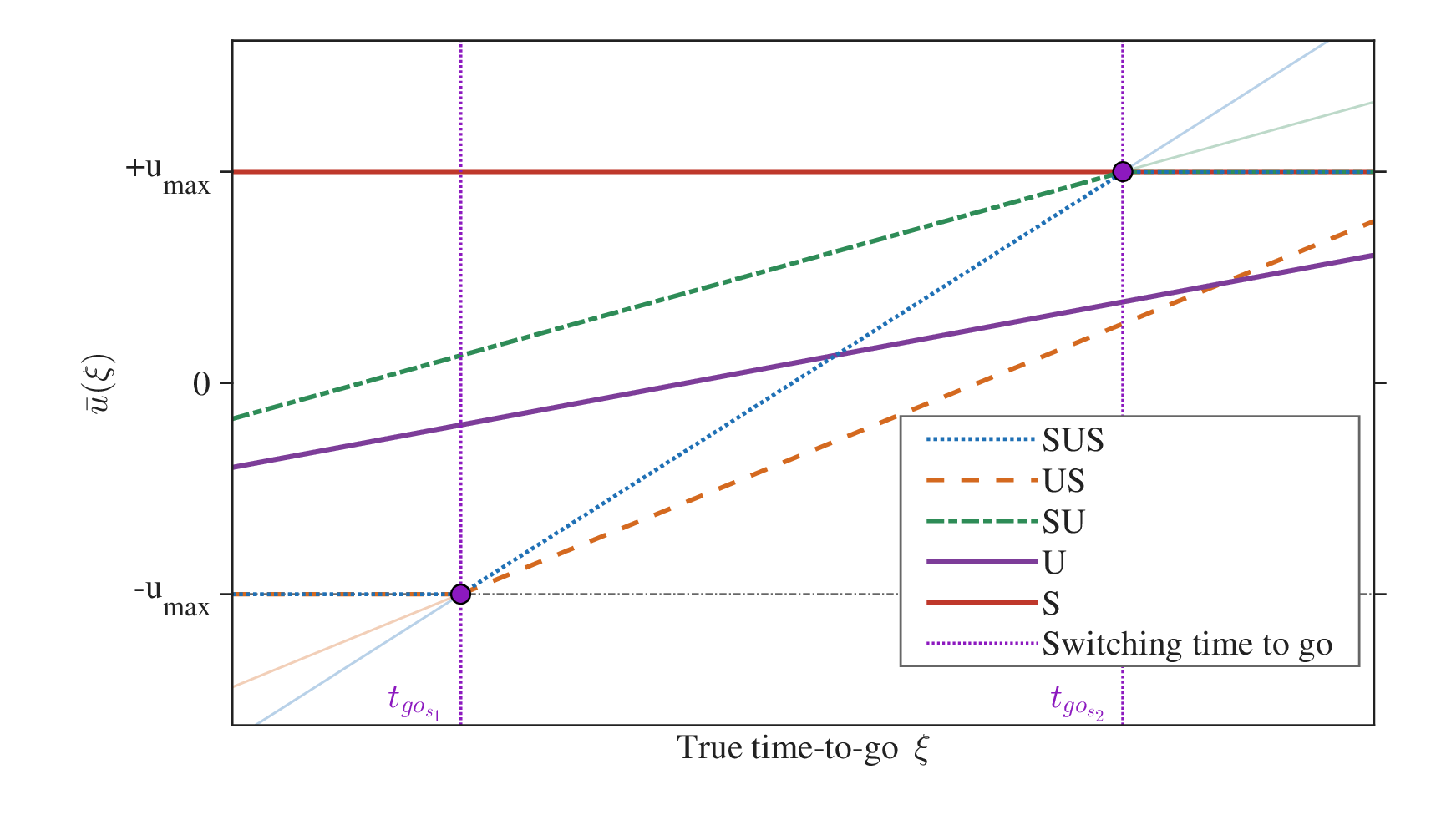}
\caption{\label{fig:saturation_mode_overlay} Command arc patterns of $\bar u$ with the switching times $t_{go_{s1}}$ and $t_{go_{s2}}$.}

\end{figure}

Figure~\ref{fig:saturation_mode_overlay} shows the typical five arc patterns $\bar u$ takes as the terminal values vary: the two-switch SUS, the single-switch US and SU, the
fully saturated S, and the fully unsaturated U.
Each name lists the arcs from the current instant forward to intercept, from left to right, S for a saturated arc and U for an unsaturated one, so in the time-to-go variable $\xi$ they appear in reverse order.
A name fixes the pattern alone; the sign $S_j$ of each
saturated arc is an unknown determined together with the switching times, and in SUS the two signs are opposite, since $\bar u_0$ reaches each bound at most once. Differentiating \eqref{eq:bounded_command_general} gives $\bar u_0'=-a\bar Z_1(t_f)\tilde B_1'$, so the direction in which $\bar u_0$ varies is set by the sign of $\bar Z_1(t_f)$. The figure shows an increasing $\bar u_0$ with time-to-go; for the opposite sign of $\bar Z_1(t_f)$ the signs of the saturated arcs are reversed.

For each of the five arc patterns, the terminal-value relations and switching conditions follow directly from the general terminal-value system \eqref{eq:Z_system_general} and switching equation \eqref{eq:switch_equation_general}, using the zero-order integrals in \eqref{eq:Q1Q2_zero} and \eqref{eq:Iij_zero}. In the fully unsaturated U mode, the terminal values reduce to the unconstrained solution in \eqref{eq:Z_tf_closed_unsat}. The resulting candidate systems are solved as described in Sec.~\ref{subsec:inner_solver}.

\subsection{Reachability of the Terminal Demands}
\label{subsec:reach_zero_order}

\subsubsection{The reachable demands}
\label{subsubsec:reachable_demands}
Perfect interception at the prescribed terminal angle corresponds to $\bar Z_1(t_f)=0$ and
$\bar Z_2(t_f)=\chi_c$. The bounded guidance law does not enforce these values but
approaches them by minimizing the terminal cost, so it is important to analyze whether the current state
admits a command within the acceleration limit that meets them exactly. Integrating the
reduced-order dynamics \eqref{eq:Zdot_reduced_main} over $[0,t_{go}]$, they are met when
the command produces the corrections
\begin{equation}
\label{eq:reach_integrals}
\int_0^{t_{go}}\tilde B_1(\xi)\,u(\xi)\,d\xi=R_1,
\quad
\int_0^{t_{go}}\tilde B_2(\xi)\,u(\xi)\,d\xi=R_2
\end{equation}
\begin{equation}
\label{eq:R1R2_def}
R_1\triangleq -Z_1(t),\quad R_2\triangleq \chi_c-Z_2(t)
\end{equation}
where $Z_1(t)$ and $Z_2(t)$ are given by \eqref{eq:Z_zero_order} and $R_1$ and $R_2$ are the required miss
 and angle corrections, respectively. For the zero-order model
$\tilde B_2=1/V_M$ is constant and $\tilde B_1(\xi)=-\xi K_M(\xi)$.

By the assumption of Sec.~\ref{subsec:zo_arcs}, $\tilde B_1$ is strictly monotonic over $[0,t_{go}]$ and does not change sign. Without loss of generality, it is taken to be strictly decreasing, so that $\tilde B_1<0$ and $\mathcal Q_1(t_{go})<0$ over $(0,t_{go}]$; the strictly increasing case is recovered by replacing $\tilde B_1$ with $-\tilde B_1$. 
The required corrections $R_1$ and $R_2$ may be of either sign.

Since $\tilde B_2$ is constant, the required angle correction fixes the integral of the command
\begin{equation}
\label{eq:mean_command_zo}
\int_0^{t_{go}} u(\xi)\,d\xi = V_M R_2
\end{equation}
Any admissible command satisfies
$\bigl|\int_0^{t_{go}} u\,d\xi\bigr|\le u_{\max} t_{go}$, so the angle correction is
reachable only if
\begin{equation}
\label{eq:angle_bound}
|R_2| \le \frac{u_{\max}\,t_{go}}{V_M}
\end{equation}
Condition \eqref{eq:angle_bound} is also sufficient for the angle correction alone. The constant command $u=V_M R_2/t_{go}$ yields the integral $V_M R_2$ required by \eqref{eq:mean_command_zo}, and its magnitude $V_M|R_2|/t_{go}$ is within the limit $u_{\max}$ when \eqref{eq:angle_bound} holds. 
It is therefore a necessary and sufficient condition for achieving $R_2$, independent of $R_1$, and when it fails, no admissible command can achieve the required angle correction.

For an $R_2$ satisfying \eqref{eq:angle_bound}, let $u_1$ and $u_2$ be admissible commands that produce $R_2$. For $0\le\theta\le1$, the command $u_\theta=(1-\theta)u_1+\theta u_2$ is also admissible, since $|u_\theta|\le(1-\theta)|u_1|+\theta|u_2|\le u_{\max}$ at every $\xi$. Moreover, $u_\theta$ produces the same $R_2$, since \eqref{eq:mean_command_zo} is linear in the command. Its miss correction

\begin{equation}
\label{eq:convex_miss}
\int_0^{t_{go}}\tilde B_1\,u_\theta\,d\xi=(1-\theta)\int_0^{t_{go}}\tilde B_1\,u_1\,d\xi+\theta\int_0^{t_{go}}\tilde B_1\,u_2\,d\xi
\end{equation}
is affine in $\theta$, and therefore takes every value between the miss corrections of $u_1$ and $u_2$. The miss corrections reachable at a fixed $R_2$ therefore form an interval, whose endpoints are identified next in Sec.~\ref{subsubsec:extreme_commands}.

\subsubsection{The extreme commands}
\label{subsubsec:extreme_commands}

Consider the two single-switch Bang-Bang controllers

\begin{equation}
\label{eq:controls_AB}
u^{+}(\xi)=
\begin{cases}
+u_{\max} & 0\le\xi\le t_{go_{s}}^{+}\\
-u_{\max} & t_{go_{s}}^{+}<\xi\le t_{go}
\end{cases}
\quad
u^{-}(\xi)=
\begin{cases}
-u_{\max} & 0\le\xi\le t_{go_{s}}^{-}\\
+u_{\max} & t_{go_{s}}^{-}<\xi\le t_{go}
\end{cases}
\end{equation}
whose switch times are fixed by the angle correction condition \eqref{eq:mean_command_zo}
\begin{equation}
\label{eq:tspm}
t_{go_{s}}^{+}=\frac{t_{go}}{2}+\frac{V_M R_2}{2 u_{\max}}, \quad
t_{go_{s}}^{-}=\frac{t_{go}}{2}-\frac{V_M R_2}{2 u_{\max}}
\end{equation}
For readability, note that the $\Box^+$ and $\Box^-$ superscripts indicate the sign of the command at 
$\xi = 0$.
Condition \eqref{eq:angle_bound} is exactly the requirement that $t_{go_{s}}^{\pm}\in[0,t_{go}]$, and the two switch times are symmetric about $t_{go}/2$, so $t_{go_{s}}^{+}+t_{go_{s}}^{-}=t_{go}$. Both commands are therefore admissible and yield $R_2$ by construction. As shown next, they produce the two endpoints of the miss-correction interval of Sec.~\ref{subsubsec:reachable_demands}.

Among all admissible commands that yield \(R_2\), \(u^{+}\) produces the largest miss correction and \(u^{-}\) the smallest. To show this, let \(u\) be any command satisfying \(|u(\xi)|\le u_{\max}\) over \([0,t_{go}]\) and yielding \(R_2\). By \eqref{eq:mean_command_zo}, its integral equals that of \(u^{+}\), and hence
\(\int_0^{t_{go}}\bigl(u^{+}-u\bigr),d\xi=0.\) Therefore, any constant multiple of this zero integral may be subtracted from the miss-correction difference without changing its value. Choosing the constant as \(\tilde B_1(t_{go_{s}}^{+})\) gives
\begin{equation}
\label{eq:shifted_integral}
\begin{aligned}
\int_0^{t_{go}}\tilde B_1\,(u^{+}-u)\,d\xi
&=\int_0^{t_{go}}\bigl[\tilde B_1(\xi)-\tilde B_1(t_{go_{s}}^{+})\bigr](u^{+}-u)\,d\xi \ge 0
\end{aligned}
\end{equation}

The product of the two factors of the integrand is nonnegative on either side of $t_{go_{s}}^{+}$. For $\xi\le t_{go_{s}}^{+}$, the command difference is $u^{+}-u=u_{\max}-u\ge 0$, while $\tilde B_1(\xi)\ge\tilde B_1(t_{go_{s}}^{+})$ because $\tilde B_1$ is strictly decreasing, so the bracket is nonnegative. For $\xi>t_{go_{s}}^{+}$, the command difference is $u^{+}-u=-u_{\max}-u\le 0$, while the bracket is negative. Thus, the product is nonnegative over the entire interval, and the right-hand side of \eqref{eq:shifted_integral} is nonnegative. Therefore, the miss correction produced by $u^{+}$ is at least as large as that produced by $u$.

Repeating the argument with the constant $\tilde B_1(t_{go_{s}}^{-})$, for $\xi\le t_{go_{s}}^{-}$ the bracket is nonnegative, while $u^{-}-u=-u_{\max}-u\le 0$. For $\xi>t_{go_{s}}^{-}$, the bracket is negative, while $u^{-}-u=u_{\max}-u\ge 0$. Thus, the product is nonpositive over the entire interval, and the miss correction produced by $u^{-}$ is no larger than that produced by $u$, \(\int_0^{t_{go}}\tilde B_1\,(u^{-}-u)\,d\xi \le 0\).

The argument relies on the monotonicity of $\tilde B_1$. Since $\tilde B_1(0)=0$, a strictly decreasing $\tilde B_1$ is nonpositive, and the differences relative to its value at the switch have the signs used above. For a strictly increasing $\tilde B_1$, the roles of $u^{+}$ and $u^{-}$ are reversed.

The endpoints of the miss correction interval are found by integrating $\tilde B_1$ over the two arcs of each command. For $u^{+}$, using $\mathcal Q_1$ in \eqref{eq:Q_def}
\begin{equation}
\label{eq:R1A_arcs}
R_1^{+}=u_{\max}\int_0^{t_{go_{s}}^{+}}\tilde B_1\,d\xi-u_{\max}\int_{t_{go_{s}}^{+}}^{t_{go}}\tilde B_1\,d\xi
=u_{\max}\bigl\{\mathcal Q_1(t_{go_{s}}^{+})-\bigl[\mathcal Q_1(t_{go})-\mathcal Q_1(t_{go_{s}}^{+})\bigr]\bigr\}
\end{equation}
Rearranging and repeating the derivation for $u^{-}$ with the arcs interchanged yields the endpoints
\begin{equation}
\label{eq:R1_bounds}
R_1^{+}=u_{\max}\bigl[2\mathcal Q_1(t_{go_{s}}^{+})-\mathcal Q_1(t_{go})\bigr],
\quad
R_1^{-}=u_{\max}\bigl[\mathcal Q_1(t_{go})-2\mathcal Q_1(t_{go_{s}}^{-})\bigr]
\end{equation}
Since $\mathcal Q_1$ is given by the polynomial in \eqref{eq:Q1Q2_zero}, both endpoints are polynomials in $t_{go_{s}}^{\pm}$ and $t_{go}$ evaluated with the fitted coefficients, and both may be of either sign. Their extreme values occur at the two extremes of the angle correction: at $t_{go_{s}}^{+}=0$ and $t_{go_{s}}^{-}=t_{go}$, both equal $-u_{\max}\mathcal Q_1(t_{go})$, and at $t_{go_{s}}^{+}=t_{go}$ and $t_{go_{s}}^{-}=0$, both equal $u_{\max}\mathcal Q_1(t_{go})$. 

Applying the result that, for every $R_2$ satisfying \eqref{eq:angle_bound}, $u^{+}$ and $u^{-}$ provide the largest and smallest miss corrections, respectively, yields $R_1^{-}\le R_1^{+}$, so the two endpoints bracket a nonempty interval. The terminal demands can therefore be met by a command within the limit when
\begin{equation}
\label{eq:reach_condition}
|R_2|\le\frac{u_{\max}\,t_{go}}{V_M}
\quad\text{and}\quad
R_1^{-}\le R_1\le R_1^{+}
\end{equation}
Together, the two conditions are necessary and sufficient. Without the first, no admissible command produces the angle correction, and without the second, the demanded $R_1$ lies outside the range of miss corrections achievable by admissible commands that produce $R_2$.
 
When \eqref{eq:angle_bound} holds with equality, the two switch times reach the endpoints of $[0,t_{go}]$, one at $0$ and the other at $t_{go}$, so each extreme command has a single arc and both reduce to $u\equiv u_{\max}\operatorname{sgn}R_2$. This is the only admissible command that produces this angle correction, so no freedom in the miss correction remains: the two endpoints coincide at $u_{\max}\operatorname{sgn}(R_2)\,\mathcal Q_1(t_{go})$, and the interval degenerates to that single value.

The miss-correction interval endpoints in \eqref{eq:R1_bounds} depend on the commanded angle through $R_2$ and the switch times \eqref{eq:tspm}, while the required miss correction $R_1=-Z_1(t)$ is fixed by the current state. The next section, Sec.~\ref{subsubsec:limiting_angle}, determines the limiting commanded angles for which this required miss correction remains within the attainable interval.

\subsubsection{The limiting commanded angles for interception}
\label{subsubsec:limiting_angle}

For each candidate value of $\chi_c$, the corresponding angle correction $R_2=\chi_c-Z_2(t)$, where $Z_2(t)$ is fixed by the state, determines the switch times through \eqref{eq:tspm}, which in turn determine the endpoints of the attainable interval $[R_1^{-},R_1^{+}]$ in \eqref{eq:R1_bounds}. 
A given commanded angle $\chi_c$ is therefore reachable while intercepting the target if and only if the fixed value $R_1$ lies in the interval \([R_1^{-}(\chi_c),R_1^{+}(\chi_c)]\). The direction in which these endpoints vary with $\chi_c$ follows by differentiating \eqref{eq:R1_bounds} along \eqref{eq:tspm}, using the chain rule:
\begin{equation}
\label{eq:dR_dchi_chain}
\frac{dR_1^{+}}{d\chi_c}=\frac{dR_1^{+}}{dt_{go_{s}}^{+}}\frac{dt_{go_{s}}^{+}}{dR_2}\frac{dR_2}{d\chi_c},
\quad
\frac{dR_1^{-}}{d\chi_c}=\frac{dR_1^{-}}{dt_{go_{s}}^{-}}\frac{dt_{go_{s}}^{-}}{dR_2}\frac{dR_2}{d\chi_c}
\end{equation}
The three factors in each derivative follow, respectively, from \eqref{eq:R1_bounds}, \eqref{eq:tspm}, and $R_2=\chi_c-Z_2(t)$. Using also \(d\mathcal Q_1(\xi)/d\xi=\tilde B_1(\xi)\) from the definition in \eqref{eq:Q_def} gives
\begin{equation}
\label{eq:dR_dchi}
\frac{dR_1^{+}}{d\chi_c}=V_M\,\tilde B_1(t_{go_{s}}^{+}),
\quad
\frac{dR_1^{-}}{d\chi_c}=V_M\,\tilde B_1(t_{go_{s}}^{-})
\end{equation}

Both derivatives are nonpositive and strictly negative whenever the corresponding switch time lies in $(0,t_{go}]$, since $\tilde B_1<0$ there. Thus, as $\chi_c$ decreases, both endpoints of the miss correction interval increase and the entire interval shifts upward, while $R_1$ remains fixed. Although both endpoints increase, as long as $R_1$ remains within the attainable interval, they lie on opposite sides of $R_1$ and therefore affect reachability differently. The upper endpoint $R_1^{+}$ moves away from $R_1$, so the condition $R_1\le R_1^{+}$ becomes easier to satisfy. In contrast, the lower endpoint $R_1^{-}$ approaches $R_1$ from below, so the condition $R_1\ge R_1^{-}$ can eventually become tight and then fail. The lower limit of $\chi_c$ is therefore the value at which $R_1^{-}=R_1$.

Whether such a value of $\chi_c$ exists is determined by the range of $R_1^{-}$. Since \eqref{eq:tspm} relates $\chi_c$ and $t_{go_{s}}^{-}$ through an affine map, with $t_{go_{s}}^{-}$ increasing from $0$ to $t_{go}$ as $\chi_c$ decreases across the angle window, it is equivalent, and simpler, to examine $R_1^{-}$ as a function of the switch time. Differentiating \eqref{eq:R1_bounds},

\begin{equation}
\label{eq:dR1B_dts}
\frac{dR_1^{-}}{dt_{go_{s}}^{-}}=-2u_{\max}\tilde B_1(t_{go_{s}}^{-})>0, \quad t_{go_{s}}^{-}\in(0,t_{go}]
\end{equation}
Thus, $R_1^{-}$ is continuous and strictly increasing. Its values at the endpoints of $[0,t_{go}]$ are the values at which the two endpoints coincide, as shown in Sec.~\ref{subsubsec:extreme_commands}: $u_{\max}\mathcal Q_1(t_{go})$ at $t_{go_{s}}^{-}=0$ and $-u_{\max}\mathcal Q_1(t_{go})$ at $t_{go_{s}}^{-}=t_{go}$. Therefore, $R_1^{-}$ spans the interval between these two values and assumes each value in the interval exactly once.

Consequently, the equation $R_1^{-}=R_1$ has a unique solution if and only if $R_1$ lies in this interval. Using $R_1=-Z_1(t)$ and $\mathcal Q_1(t_{go})<0$, this condition becomes

\begin{equation}
\label{eq:miss_authority}
|Z_1(t)| \le u_{\max}\,\bigl|\mathcal Q_1(t_{go})\bigr|
\end{equation}
The two extreme values attained by $R_1^{-}$ also define the range of miss corrections available when there is no angle demand. Since $\tilde B_1$ does not change sign, these extremes are produced by the constant commands

\begin{equation}
\label{eq:miss_extremes}
\int_0^{t_{go}}\tilde B_1\,(\mp u_{\max})\,d\xi=\mp u_{\max}\,\mathcal Q_1(t_{go})
\end{equation}
which are the limiting forms of $u^{-}$ at the two ends of the angle window. Condition \eqref{eq:miss_authority} therefore states that the required miss correction $R_1$ alone lies within the available control authority. If it fails, no admissible command can produce $R_1$, regardless of the commanded angle.

When \eqref{eq:miss_authority} holds strictly, the equation $R_1^{-}=R_1$ has a root strictly inside $(0,t_{go})$. Hence, as $\chi_c$ decreases, the angle-specific reachability condition $R_1\ge R_1^{-}(\chi_c)$ becomes tight before the angle bound \eqref{eq:angle_bound} is reached. The limiting switch time $t_{go_{s}}^{\star}$ is therefore determined by $R_1^{-}=R_1$, which, using \eqref{eq:R1_bounds} and $R_1=-Z_1(t)$, reads

\begin{equation}
\label{eq:root_eq}
\mathcal Q_1(t_{go_{s}}^{\star})-\left[\frac{\mathcal Q_1(t_{go})}{2}+\frac{Z_1(t)}{2 \, u_{\max}}\right]=0
\end{equation}
a polynomial equation in \(t_{go_{s}}^{\star}\) of similar form to the switching conditions, whose root is sought in \([0,t_{go}]\). Rearranging \eqref{eq:tspm} with \(t_{go_{s}}^{-}=t_{go_{s}}^{\star}\) and substituting \(R_2=\chi_c-Z_2(t)\) then gives the minimal commanded angle

\begin{equation}
\label{eq:chi_min}
\chi_c^{\min}=Z_2(t)+\frac{u_{\max}}{V_M}\left(t_{go}-2\,t_{go_{s}}^{\star}\right)
\end{equation}
Both remaining reachability conditions are met at $\chi_c^{\min}$. The angle bound \eqref{eq:angle_bound} holds because $t_{go_{s}}^{\star}\in[0,t_{go}]$, and the upper condition holds because $R_1=R_1^{-}\le R_1^{+}$.

Similarly, as $\chi_c$ increases, both endpoints of the miss correction interval decrease and the entire interval shifts downward by \eqref{eq:dR_dchi}, so the upper endpoint descends toward $R_1$ and the condition $R_1\le R_1^{+}$ is the one that fails first, setting the upper end of the reachable range. Repeating the argument with $R_1^{+}$ in place of $R_1^{-}$ gives the upper limiting switch time $t_{go_{s}}^{\star\star}$, at which $R_1^{+}=R_1$. This switch time exists and is unique under the same condition \eqref{eq:miss_authority}, since $R_1^{+}$ sweeps the same range as $R_1^{-}$ over $[0,t_{go}]$. The corresponding maximal commanded angle is then

\begin{equation}
\label{eq:root_eq_upper}
\chi_c^{\max}=Z_2(t)+\frac{u_{\max}}{V_M}\left(2\,t_{go_{s}}^{\star\star}-t_{go}\right), \quad  
\mathcal Q_1(t_{go_{s}}^{\star\star})-\left[\frac{\mathcal Q_1(t_{go})}{2}-\frac{Z_1(t)}{2\,u_{\max}}\right]=0
\end{equation}
Each condition in \eqref{eq:reach_condition} holds on an interval of $\chi_c$, so the reachable commanded angles form the interval $\bigl[\chi_c^{\min},\chi_c^{\max}\bigr]$.

For higher-order missile dynamics, $\tilde B_2$ is no longer constant, so the angle correction no longer fixes the switch times of the extreme commands through the command integral alone. The interval structure of the reachable miss corrections survives, since it follows from linearity alone, but the endpoints must be found jointly with the angle correction, making the analysis substantially more complicated.

\section{Implementation}
\label{sec:algorithm_implementation}

This section describes the numerical implementation of the proposed Intercept-Angle Bounded Optimal Guidance Law (IABOGL) in a nonlinear engagement, together with the unconstrained Intercept-Angle Optimal Guidance Law (IAOGL) of Sec.~\ref{subsec:unconstrained_command} and the OGL-CTIA \cite{shaferman_linear_2008}. Both are used to initialize the IABOGL, as described in Sec.~\ref{subsec:geometry_and_tgo}, and both are compared against it in Sec.~\ref{sec:simulation_results}. At each guidance step, the geometric coefficients and time-to-go are refined over the remaining flight time, the current projected variables are evaluated, and the inner solver determines the bounded command.
The following subsections describe these elements in turn and then present the complete algorithm.

\subsection{Initialization, Geometry Coefficients, and Time-to-Go}
\label{subsec:geometry_and_tgo}

The bounded solution at the first guidance step is initialized in two simulation passes. The first is an unbounded pre-run of the OGL-CTIA \cite{shaferman_linear_2008}, which holds \(K_M\) and \(K_T\) constant at each guidance step; their realized histories along that run are sampled and fitted by least squares \cite{mendel_lessons_1995} to the cubic approximations \eqref{eq:K_cubic_app}. Since \(K_M\) and \(K_T\) are cosines and cannot exceed unity, the fitted coefficients are rescaled whenever either fitted cubic exceeds unity. The second pass repeats the run with the unbounded IAOGL law on those cubics, solving at each step for the time-to-go 
\begin{equation}
\label{eq:range_relation_tgo}
\int_0^{t_{go}} \bigl[V_M K_M(\xi)+V_T K_T(\xi)\bigr]\,d\xi = r
\end{equation}
which is a polynomial equation in \(t_{go}\). Its roots are computed using the MATLAB \texttt{roots} function, and the positive real root nearest the previous estimate is selected. When no such root exists, the kinematic estimate
\begin{equation}
\label{eq:tgo_guess}
t_{go}^{\mathrm{kin}}=\frac{r}{V_c},\quad V_c=V_M \cos(\gamma_M-\sigma)+V_T \cos(\gamma_T+\sigma)
\end{equation}
is used as a fallback.
The functions \(K_M(\xi)\) and \(K_T(\xi)\) are sampled along this trajectory and refitted by least squares again.
The bounded solve then starts from these estimates, together with the predicted terminal values of the IAOGL, and generates an updated bounded solution.

At each bounded guidance step (including the first with the initialization described above), the geometry is updated from a trajectory predicted over \([0,t_{go}]\) using the current open-loop bounded solution, with its arc sequence, signs, and terminal values held fixed. The functions \(K_M(\xi)\) and \(K_T(\xi)\) are sampled along this trajectory and refitted by least squares. The time-to-go is the value consistent with the fitted geometry, obtained from the range relation in \eqref{eq:range_relation_tgo}.

\subsection{Approximation of the Current Projected Variables}
\label{subsec:current_Z_approximation}

The \(\bar Z_1(t_f)\) and \(\bar Z_2(t_f)\) terminal-value and switching formulas require the current projected quantities \(Z_1(t)\) and \(Z_2(t)\), given by \eqref{eq:Z_zero_order}. The target contribution is already in closed polynomial form through the fitted geometry coefficients \(k_{T,i}\). The remaining quantity, \(y+\dot y\,t_{go}\), is not directly measured or estimated in either the nonlinear simulation or a real engagement. Under the small-deviation approximation from the collision triangle, \(\dot y\) is the transverse relative velocity, while the current lateral displacement is zero because we re-linearize at every time step, therefore,
\begin{equation}
\label{eq:zem_small_dev}
y+\dot y\,t_{go}\approx v_\theta\,t_{go},
\quad
v_\theta=V_T\sin(\gamma_T+\sigma)-V_M\sin(\gamma_M-\sigma)
\end{equation}
From the relative transverse kinematics \eqref{eq:sigmadot}, \(v_\theta=r\,\dot\sigma\), so
the implemented miss-related projected variable is
\begin{equation}
\label{eq:Z1_hat_impl}
Z_1(t)=\dot\sigma\,r\,t_{go}+\sum_{i=0}^{3}\frac{k_{T,i}\,a_T}{i+2}\,t_{go}^{\,i+2}
\end{equation}
The \(\dot\sigma\)-based form is used because \(\dot\sigma\) and \(r\) can be measured or estimated directly from the seeker, and this form does not require the typical approximation \(r\approx V_c \, t_{go}\), which is less justifiable when the closing speed varies. The angle-related projected variable requires no additional approximation, since \(\Delta\gamma\) and \(a_T\) are available from the state vector in the simulation and can be estimated in a real engagement.

\subsection{Inner Switch-Times and Saturation-Mode Solver}
\label{subsec:inner_solver}

\subsubsection{The Candidate System and Its Solution}
\label{subsubsec:candidate_system}

For fixed geometry coefficients and a fixed \(t_{go}\), the inner solver determines the switching times, the arc sequence, and the terminal values \(\bar Z_1(t_f)\) and \(\bar Z_2(t_f)\). These quantities are coupled. The unsaturated command \(\bar u_0\) is built from the terminal values \eqref{eq:bounded_command_general}; the points where \(\bar u_0\) crosses \(\pm u_{\max}\) set the switching times \eqref{eq:switch_equation_general}; and the switching times, through the saturated arcs they define, fix the terminal values \eqref{eq:Z_system_general}. None can be found independently of the others, so for a given arc sequence the solver treats them as a single system and solves for them jointly.

The arc sequence itself, however, is not known in advance. Each admissible sequence therefore forms its own system, and a sequence with \(m\) switches gives a system of \(2+m\) equations, so different candidates give systems of different size. Which of them are actually solved, and how one is selected, is described below. The solver uses the dynamics only through \(\tilde B_1\), \(\tilde B_2\), and their integrals, so the same procedure applies to any missile order. Under zero-order dynamics, the sequences are the saturation modes of Sec.~\ref{sec:influence_and_integrals} and reduce to the five forms of Fig.~\ref{fig:saturation_mode_overlay}, so \(m_{\max}=2\), while higher-order dynamics may admit more.

The solver works with a candidate arc sequence: a number of switches \(m\) and a sign
\(S_j\in\{-1,+1\}\) for each saturated arc. For a given candidate, the unknowns are the two
terminal values and the \(m\) switching times
\begin{equation}
\label{eq:unknown_vector}
\mathbf y=\bigl[\,\bar Z_1(t_f),\ \bar Z_2(t_f),\ t_{go_{s}}^{T}\,\bigr]^{T}, \quad t_{go_{s}}^{T}=\bigl[\,t_{go_{s1}},\dots,t_{go_{sm}}\,\bigr],
\quad 0<t_{go_{s1}}<\dots<t_{go_{sm}}\le t_{go}
\end{equation}
and they satisfy a system \(\mathbf F(\mathbf y)=\mathbf 0\) of \(2+m\) equations: the two
terminal-value relations and the \(m\) switching conditions
\begin{subequations}
\label{eq:Fsystem}
\begin{align}
E_1&=\bigl[1+a\bar{\mathcal I}_{11}\bigr]\bar Z_1(t_f)+b\bar{\mathcal I}_{12}\bigl[\bar Z_2(t_f)-\chi_c\bigr]-Z_1(t)-\mathcal S_1 \label{eq:E1}\\
E_2&=a\bar{\mathcal I}_{12}\,\bar Z_1(t_f)+\bigl[1+b\bar{\mathcal I}_{22}\bigr]\bigl[\bar Z_2(t_f)-\chi_c\bigr]-\bigl[Z_2(t)-\chi_c\bigr]-\mathcal S_2 \label{eq:E2}\\
E_{2+j}&=\bar u_0(t_{go_{sj}})-S_j\,u_{\max},\quad j=1,\dots,m \label{eq:Eswitch}
\end{align}
\end{subequations}
The first two are the terminal-value system \eqref{eq:Z_system_general} with all terms
moved to one side; the remaining \(m\) require \(\bar u_0\) to reach the bound at each
switching time \eqref{eq:switch_equation_general}. The integrals \(\bar{\mathcal I}_{ij}\)
\eqref{eq:Ihat_defs} and the saturated contributions \(\mathcal S_1,\mathcal S_2\)
\eqref{eq:S_def} depend on the switching times, while \(\bar u_0\) depends on the terminal
values, so the unknowns cannot be determined independently and the system is solved as a whole.

The last inequality in \eqref{eq:unknown_vector} becomes non-strict only when the final arc, saturated or unsaturated, vanishes. This change in the structure of the equation set is handled by the complementarity condition of Sec.~\ref{subsubsec:mode_transitions}.

The system \eqref{eq:Fsystem} is solved by the damped Newton method \cite{facchinei_finite-dimensional_2003} (Sec.~8.1).
At iterate \(\mathbf y^{(n)}\) the Newton step \(\Delta\mathbf y^{(n)}\) solves
\begin{equation}
\label{eq:newton_step}
\mathbf J(\mathbf y^{(n)})\,\Delta\mathbf y^{(n)}=-\mathbf F(\mathbf y^{(n)})
\end{equation}
where \(\mathbf J=\partial\mathbf F/\partial\mathbf y\) is the Jacobian, given in closed
form in \ref{app:jacobian}. The iterate is updated with a step length \(\eta_n\in(0,1]\)
\begin{equation}
\label{eq:newton_update}
\mathbf y^{(n+1)}=\mathbf y^{(n)}+\eta_n\,\Delta\mathbf y^{(n)}
\end{equation}
The step length is set by backtracking: \(\eta_n\) is initialized to one and repeatedly halved until the updated switching times remain in increasing order within \([0,t_{go}]\) and \(\lVert\mathbf F\rVert_\infty\) decreases. The iteration stops when \(\lVert\mathbf F\rVert_\infty\) falls below the tolerance \(\epsilon_F\).

The remaining task is to select a candidate arc sequence and construct its initial vector $\mathbf y^{(0)}$. A candidate is accepted only if the Newton iteration converges and the resulting command passes the arcwise admissibility check in \eqref{eq:admissible}.

At the first guidance step, the terminal-value initial guess is supplied by the second initialization pass of Sec.~\ref{subsec:geometry_and_tgo}. The real roots in $(0,t_{go})$ of
\begin{equation}
\label{eq:root_init}
\bar u_0(\xi)-u_{\max}=0,
\qquad
\bar u_0(\xi)+u_{\max}=0
\end{equation}
determine the initial switch count and switching times, while the terminal-arc type is read from \(\lvert\bar u_0(0)\rvert\) and the saturation signs are taken from \(\bar u_0\) at the arc midpoints. The terminal-arc type specifies whether the arc adjacent to the intercept, at \(\xi=0\), is saturated or unsaturated.

At each later guidance step, the terminal values from the previous converged solution are used to construct \(\bar u_0\) under the current geometry. The roots of \eqref{eq:root_init} then provide the nominal number of switches \(m\) and the switching-time guesses, the terminal-arc type is read from \(\lvert\bar u_0(0)\rvert\), and the saturation signs are taken from \(\bar u_0\) at the arc midpoints. Because the arc types alternate, the terminal-arc type together with the switch count \(m\) determines the complete arc sequence. This nominal candidate is solved first and returned if accepted. Since changes in the arc structure occur only at isolated mode transitions, the nominal candidate suffices at most guidance steps.

If the nominal candidate is rejected, the solver retains the carried terminal values and varies only the arc structure. When \(m=0\), it tests the other zero-switch sequence and the two one-switch sequences. When \(m\geq1\), it tests both terminal-arc types with \(m-1\) and \(m+1\) switches. A switching time is removed by dropping the one nearest \(t_{go}\) or added midway between the last switch and \(t_{go}\); when \(m=0\), the added switch is placed at \(t_{go}/2\). The saturation signs are set from \(\bar u_0\) at the corresponding arc midpoints, and the first accepted candidate is returned.

If the first fallback fails, the carried terminal values and all quantities derived from them are discarded. A fresh terminal pair is obtained from the closed-form unconstrained solution in \eqref{eq:Z_tf_closed_unsat}, and \(\bar u_0\) is reconstructed at the current geometry. Its terminal-arc type is determined from \(\lvert\bar u_0(0)\rvert\), the roots of \eqref{eq:root_init} give the switching times and \(m\), and the saturation signs are obtained at the arc midpoints. The damped Newton method is first applied to the candidate with this terminal-arc type and \(m\) switches and, if necessary, to the candidates with both terminal-arc types and \(m-1\) or \(m+1\) switches. This second fallback is extremely rare and is retained for algorithmic robustness. In all tested cases, this procedure resulted in a converged solution.

\subsubsection{Mode Transitions}
\label{subsubsec:mode_transitions}

Over an engagement, the selected arc sequence changes as saturated arcs appear and vanish,
for example \(\mathrm{SUS}\to\mathrm{US}\to\mathrm{S}\) as intercept approaches. The arc
that vanishes is always the one at the current instant, \(\xi=t_{go}\), since that is the
active arc and \(t_{go}\) is shrinking, whereas the arc adjacent to intercept persists.
The active arc vanishes when its switching time \(t_{go_{sm}}\) reaches \(t_{go}\). At this transition, the last switching condition, \(\bar u_0(t_{go_{sm}})=S_m u_{\max}\), still holds, but the strict ordering \(t_{go_{sm}}<t_{go}\) becomes an equality. Beyond the transition, the
\(m\)-switch candidate cannot satisfy both conditions, since maintaining the switching
condition at \(t_{go_{sm}}=t_{go}\) would require the command to remain on the bound at
\(\xi=t_{go}\), whereas it has already left it. The transition therefore occurs precisely
where the \(m\)- and \((m-1\))-switch descriptions coincide, and it is this coincidence that
the complementarity condition introduced next exploits to carry the candidate across the
transition rather than immediately discard it.

To allow a candidate to continue through its own boundary, the last switching condition
\(E_{2+m}\) is replaced, for every candidate with \(m\ge1\), by a complementarity condition.
Let \(p\) denote the normalized length of the arc at \(\xi=t_{go}\) and \(q\) the normalized
margin by which the arc satisfies its corresponding condition in \eqref{eq:admissible} at the last switching time
\begin{equation}
\label{eq:pq_def}
p=\frac{t_{go}-t_{go_{sm}}}{t_{go}},\qquad
q=\begin{cases}
\dfrac{u_{\max}-S_m\,\bar u_0(t_{go_{sm}})}{u_{\max}}, & \text{arc } m \text{ unsaturated}\\[2ex]
\dfrac{S_m\,\bar u_0(t_{go_{sm}})-u_{\max}}{u_{\max}}, & \text{arc } m \text{ saturated}
\end{cases}
\end{equation}
so that \(q\ge0\) exactly when arc \(m\) can extend to \(t_{go}\). Either arc type can
vanish: under zero-order dynamics, SUS transitions to US when its saturated arc vanishes,
and US then transitions to S when its unsaturated arc vanishes.

The saturated arc adjacent to \(\xi=t_{go}\) is present when \(p>0\) and \(q=0\), with the command at \(t_{go_{s_m}}\) sitting on the bound, and absent when \(p=0\) and \(q>0\), with the command having left the bound.
These are the complementarity conditions, which are enforced
through the Fischer--Burmeister function \cite{fischer_special_1992}, written here for two nonnegative quantities 

\begin{equation}
\label{eq:phi}
\varphi(p,q)=p+q-\sqrt{p^2+q^2}
\end{equation}

for which \(\varphi(p,q)=0\) holds if and only if \(p\ge0\), \(q\ge0\), and \(pq=0\). It is
used in place of the product condition \(pq=0\) because it is smooth away from the origin
and encodes the sign constraints together with the product condition in a single equation,
allowing the Newton iteration with a closed-form Jacobian to be applied directly away from the
transition point; in contrast, \(pq=0\) alone admits \(p<0\) or \(q<0\), whereas the direct
condition \(\min(p,q)=0\) is nonsmooth.

The condition \(\varphi=0\) is evaluated at each Newton iteration in place of \(E_{2+m}\),
not only at a transition. While the arc has positive length, \(p>0\) forces \(q=0\),
recovering the ordinary switching condition; as the arc vanishes, \(p\to0\) allows \(q>0\),
and the switching time reaches \(t_{go}\), the end of the interval. The \(m\)-switch
candidate therefore remains solvable with a zero-length final arc and converges to the same
solution as the candidate with that arc removed, so the selection at the boundary does not
affect the command.

\subsection{Algorithm Summary}
\label{subsec:algorithm_summary}
The complete guidance algorithm is given in two parts. Algorithm~\ref{alg:bogl_main} is the
guidance loop and Algorithm~\ref{alg:inner_solver} is the inner solver of Sec.~\ref{subsec:inner_solver}, which returns the terminal values and switching times for a fixed geometry and time-to-go.

\begin{algorithm}[!htbp]
\caption{IABOGL Guidance in a Nonlinear Engagement}
\label{alg:bogl_main}
\begin{algorithmic}[1]
\State Run the unbounded OGL-CTIA pre-run with constant \(K_M\), \(K_T\) and the kinematic
       \(t_{go}\) (\eqref{eq:u0_general}, \eqref{eq:Z_tf_closed_unsat}, \eqref{eq:tgo_guess})
\State Fit the cubics \(K_M\), \(K_T\) (\eqref{eq:K_cubic_app}) with \(t_{go}\) set to the flight time of the OGL-CTIA pre-run
\State Run the unbounded IAOGL pre-run on \(K_M\), \(K_T\) and the  \(t_{go}\) of \eqref{eq:range_relation_tgo}; set
       \(\bar Z_1(t_f)\), \(\bar Z_2(t_f)\), the arc sequence, and
       \(t_{go_{s}}\) from its terminal values and from the crossings of
       its command with \(\pm u_{\max}\)
       \State Refit the cubics \(K_M\), \(K_T\) (\eqref{eq:K_cubic_app}) with \(t_{go}\) set to the flight time of the IAOGL pre-run
      \State Set: \(\text{First step flag}=1\)
\For{each guidance step}
    \State Compute \(Z_1(t)\), \(Z_2(t)\) on the current \(K_T\) and \(t_{go}\)
           (\eqref{eq:Z1_hat_impl}, \eqref{eq:Z_zero_order}, \eqref{eq:range_relation_tgo})
    \State Solve for \(\bar Z_1(t_f)\), \(\bar Z_2(t_f)\), \(m\), \(t_{go_s}\), and \(S\)
            (Algorithm~\ref{alg:inner_solver})
    \For{refinement pass \(i=1,\dots,(1+\text{First step flag})\)}
        \State Predict the trajectory over \([0,t_{go}]\) under the current solution and
               refit \(K_M\), \(K_T\) (\eqref{eq:K_cubic_app})
        \State Solve the range relation for \(t_{go}\) (\eqref{eq:range_relation_tgo}),
               fallback (\eqref{eq:tgo_guess})
        \State Compute \(Z_1(t)\), \(Z_2(t)\)
               (\eqref{eq:Z1_hat_impl}, \eqref{eq:Z_zero_order})
               \State Solve for \(\bar Z_1(t_f)\), \(\bar Z_2(t_f)\), \(m\), \(t_{go_s}\), and \(S\)
            (Algorithm~\ref{alg:inner_solver})
    \EndFor
    \State Apply \(\bar u(t_{go})\) (\eqref{eq:bounded_command_general})
    \State Set: \(\text{First step flag}=0\)
\EndFor
\end{algorithmic}
\end{algorithm}

\begin{algorithm}[!t]
\caption{Inner Solver for the Terminal Values and Switching Times}
\label{alg:inner_solver}
\begin{algorithmic}[1]
\Require \(t_{go}\), \(K_M\), \(Z_1\), \(Z_2\), \(\bar Z_1(t_f)\), \(\bar Z_2(t_f)\) 
\State \label{alg:build}Build \(\bar u_0\) from \(\bar Z_1(t_f)\), \(\bar Z_2(t_f)\), \(K_M\), and \(t_{go}\) 
(\eqref{eq:bounded_command_general})
\State Set the terminal-arc type to saturated if
       \(\lvert\bar u_0(0)\rvert>u_{\max}\), and to unsaturated otherwise
\State Set \(t_{go_{s}}\) to the roots of \eqref{eq:root_init} in
       \((0,t_{go})\), and set \(m\) to their number
\State Take the sign of each saturated arc from \(\bar u_0\) at that
       arc's midpoint
\State Solve \eqref{eq:Fsystem} by damped Newton
       (\eqref{eq:newton_step}). For \(m\ge1\), \(E_{2+m}\) in \eqref{eq:Fsystem} is replaced  by \(\varphi=0\) (\eqref{eq:phi}), with \(p\) and \(q\) from \eqref{eq:pq_def}
\If{\(\|\mathbf F\|_\infty<\epsilon_F\) and \eqref{eq:admissible} holds}
    \State \Return \(\bar Z_1(t_f)\), \(\bar Z_2(t_f)\), \(m\), \(S\), \(t_{go_s}\)
\EndIf

\If{$m=0$}
\State Form the other zero-switch and the two one-switch
candidates, with the one-switch time initialized at $t_{go}/2$
\Else
\State Form candidates with both terminal-arc types and $m-1$ switches
by dropping the switching time nearest $t_{go}$
\State Form candidates with both terminal-arc types and $m+1$ switches
by inserting a switching time midway between the last switch and $t_{go}$
\EndIf

\For{each candidate}
    \State Take the sign of each saturated arc from $\bar u_0$ at that
    arc's midpoint
\State Solve \eqref{eq:Fsystem} by damped Newton
       (\eqref{eq:newton_step}). 
       For \(m\ge1\), \(E_{2+m}\) in \eqref{eq:Fsystem} is replaced  by \(\varphi=0\) (\eqref{eq:phi}), with \(p\) and \(q\) from \eqref{eq:pq_def} 
    \If{$\lVert\mathbf F\rVert_\infty<\epsilon_F$ and
    \eqref{eq:admissible} holds}
        \State \Return the accepted candidate's $\bar Z_1(t_f)$,
        $\bar Z_2(t_f)$, $m$, $S$, $t_{go_s}$
    \EndIf
\EndFor  \label{alg:lastcand}

\State Replace $\bar Z_1(t_f)$ and $\bar Z_2(t_f)$ with the
unconstrained solution in \eqsref{eq:Z_tf_closed_unsat}

\State Repeat lines \ref{alg:build}--\ref{alg:lastcand} with these terminal values
\State \Return no accepted candidate

\end{algorithmic}
\end{algorithm}

\section{Simulation Results}
\label{sec:simulation_results}
This section presents the nonlinear simulation results of the proposed bounded optimal guidance law. The engagement begins with a head-on geometry, and Table~\ref{tab:sim_params} lists the nominal simulation parameters. The missile is modeled with ideal dynamics, and the target performs a constant \(+5g\) maneuver throughout the engagement. The nonlinear kinematics of Sec.~\ref{subsec:geometry_kinematics} are integrated using the MATLAB \texttt{ode45} solver, with the maximum step reduced from \(10^{-2}\)\,s to \(10^{-5}\)\,s once the range falls below 100 m to improve the accuracy of the terminal quantities. The guidance law is evaluated at 100 Hz, and its command is held constant between updates; inside the blind range, the last command is held. Each run terminates at closest approach, identified by the first sign reversal of the closing speed, and the miss distance is the minimum range attained. The terminal-angle error is \((\gamma_T+\gamma_M)-\chi_c\), evaluated at that instant.

\begin{table*}[!htbp]
\caption{\label{tab:sim_params} Simulation parameters for the representative engagement.}
\centering
\begin{tabular}{lcc@{\hspace{1.0em}}lcc}
\hline\hline
Parameter & Symbol & Value & Parameter & Symbol & Value \\
\hline
Missile initial position & \((x_M(0),y_M(0))\) & \((0,0)\) m
  & Target acceleration & \(a_T\) & \(+5g\) \\
Target initial position & \((x_T(0),y_T(0))\) & \((2000,0)\) m
  & Missile acceleration limit & \(u_{\max}\) & \(40g\) \\
Missile speed & \(V_M\) & 500 m/s
  & Miss-distance weight & \(a\) & \(10^5\) \\
Target speed & \(V_T\) & 300 m/s
  & Terminal-angle weight & \(b\) & \(10^8\) \\
Initial missile flight-path angle & \(\gamma_M(0)\) & \(0^\circ\)
  & Blind range & \(R_{\mathrm{blind}}\) & 50 m \\
Initial target flight-path angle & \(\gamma_T(0)\) & \(0^\circ\)
  & Commanded terminal angle & \(\chi_c\) & \(-6^\circ\) \\
\hline\hline
\end{tabular}
\end{table*}

\begin{figure}[!htbp]
\centering
\includegraphics[width=0.65\textwidth]{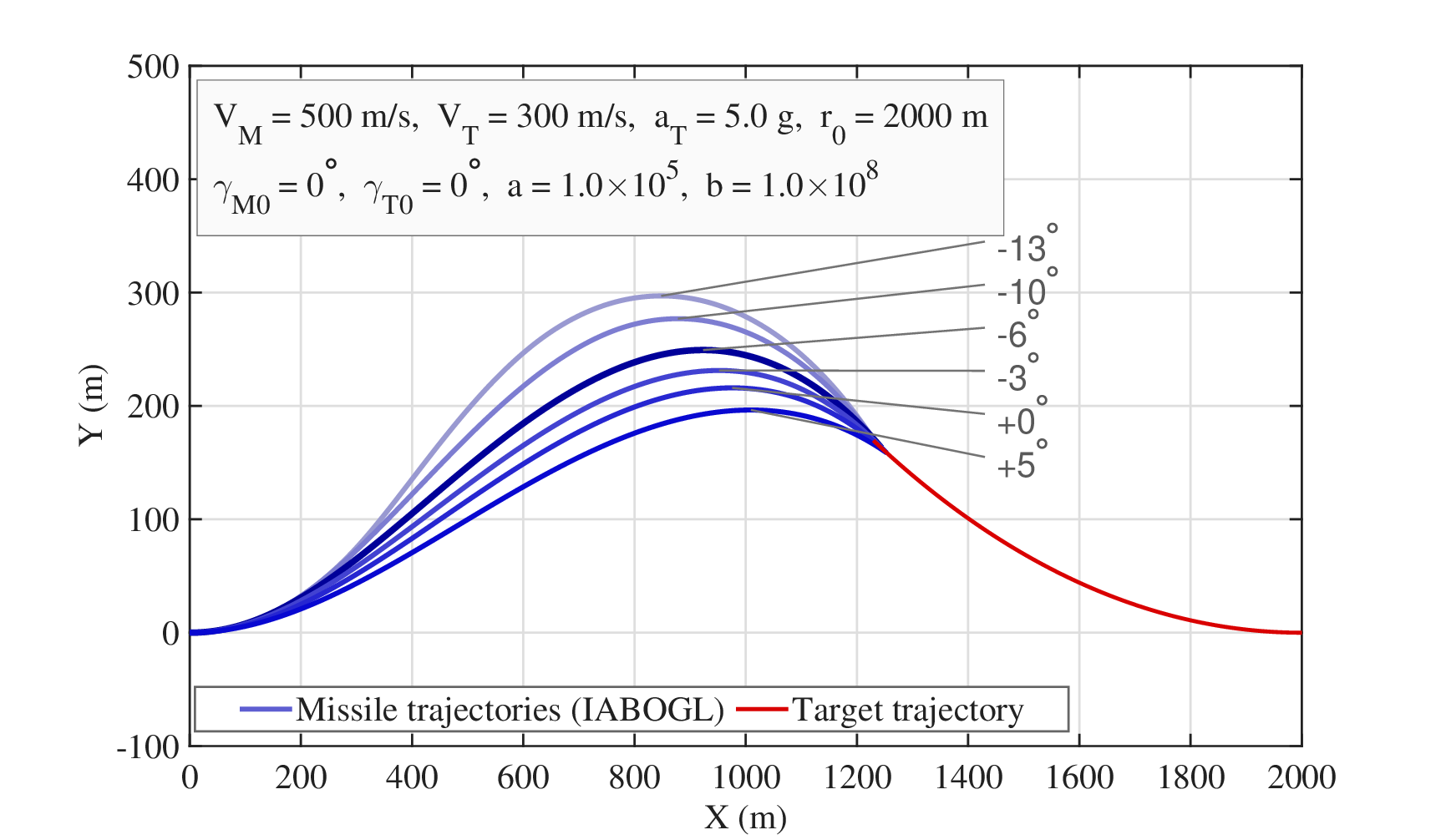}
\caption{\label{fig:show_cases_trajectories} Intercept-angle sweep - Trajectories.}
\end{figure}

\begin{figure}[!htbp]
\centering
\includegraphics[width=0.65\textwidth]{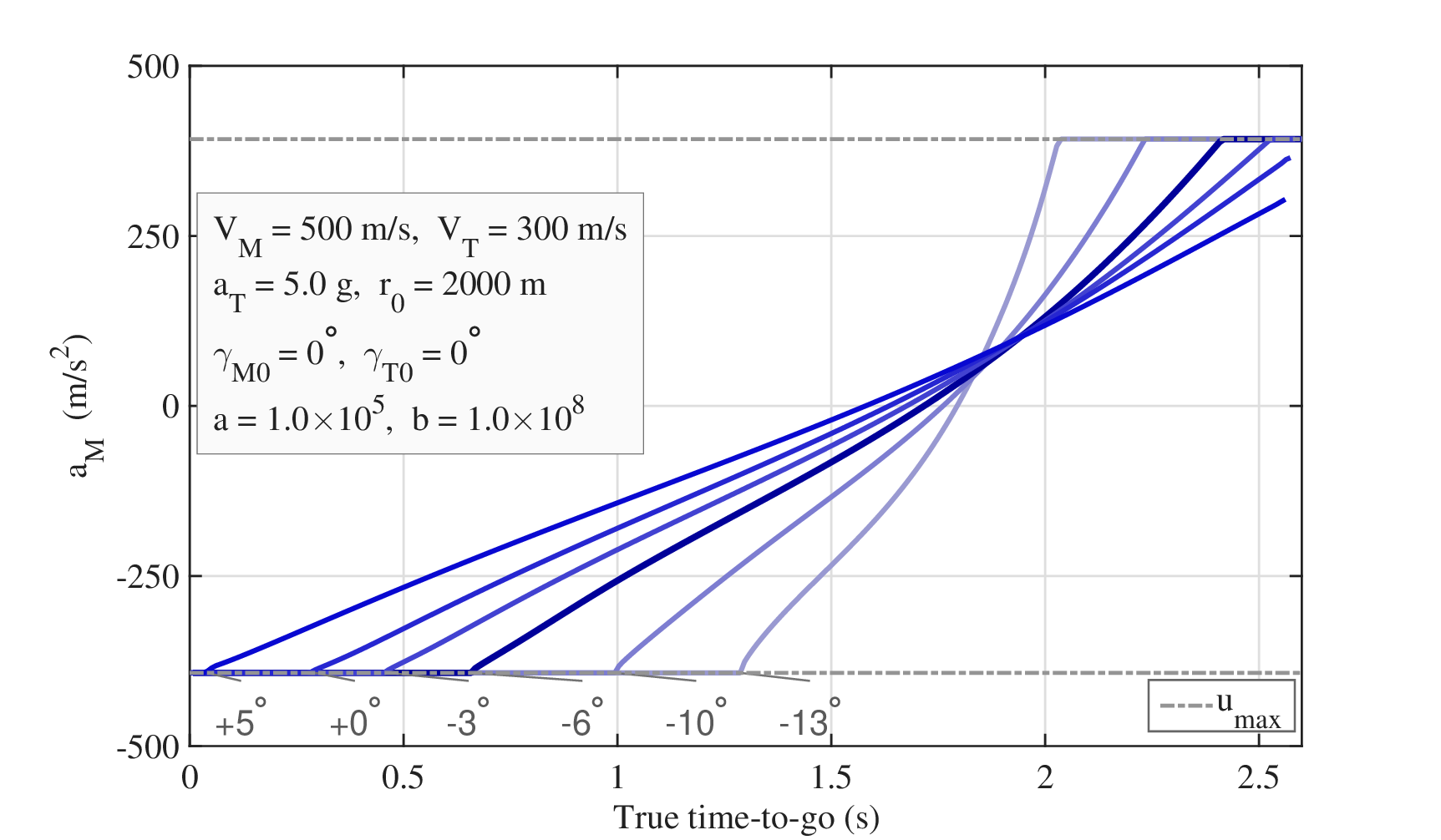}
\caption{\label{fig:show_case_commands} Intercept-angle sweep - Acceleration commands.}
\end{figure}

Figures~\ref{fig:show_cases_trajectories} and \ref{fig:show_case_commands} present an intercept-angle sweep and illustrate how the bounded guidance law reshapes the interception under the same initial engagement geometry. Although all cases intercept the target, the required lateral displacement changes substantially with \(\chi_c\). More demanding terminal-angle commands, especially on the negative side, require the missile to depart farther from the nominal collision path and to establish a steeper approach near intercept.

The command histories in Fig.~\ref{fig:show_case_commands} reflect the same trend. More demanding negative values of \(\chi_c\) require earlier and larger commands, with longer intervals on the acceleration bound, whereas larger positive angles are achieved with milder profiles. To examine how this behavior develops over time, the following analysis focuses on a representative run with \(\chi_c=-6^\circ\).

\subsection{Analysis of a Representative Sample Run}

Table~\ref{tab:case_m6_summary} summarizes the terminal miss distance and terminal-angle error for the representative case with \(\chi_c=-6^\circ\). The IABOGL is compared with the IAOGL, which uses the same time-varying \(K_M(\xi)\) and \(K_T(\xi)\) and the same time-to-go but does not account for the acceleration bound in its derivation, and with the constrained terminal intercept angle optimal guidance law (OGL-CTIA) of \cite{shaferman_linear_2008}, which likewise does not account for the bound but holds \(K_M\) and \(K_T\) constant and uses the kinematic time-to-go. Both comparison laws are saturated at \(u_{\max}\). The IABOGL yields substantially smaller terminal errors than either law, even when significant saturation occurs near the end of the engagement.

\begin{table}[!htbp]
\caption{\label{tab:case_m6_summary} Terminal performance for the representative run at \(\chi_c=-6^\circ\)}
\centering
\begin{tabular}{lcc}
\hline\hline
Guidance law & Miss distance (m) & Terminal-angle error (deg) \\
\hline
IABOGL   & 0.06 & 0.10 \\
IAOGL    & 4.63  & 3.73 \\
OGL-CTIA & 5.68  & 5.31 \\
\hline\hline
\end{tabular}
\end{table}

\begin{figure}[!htbp]
\centering
\includegraphics[width=0.65\textwidth]{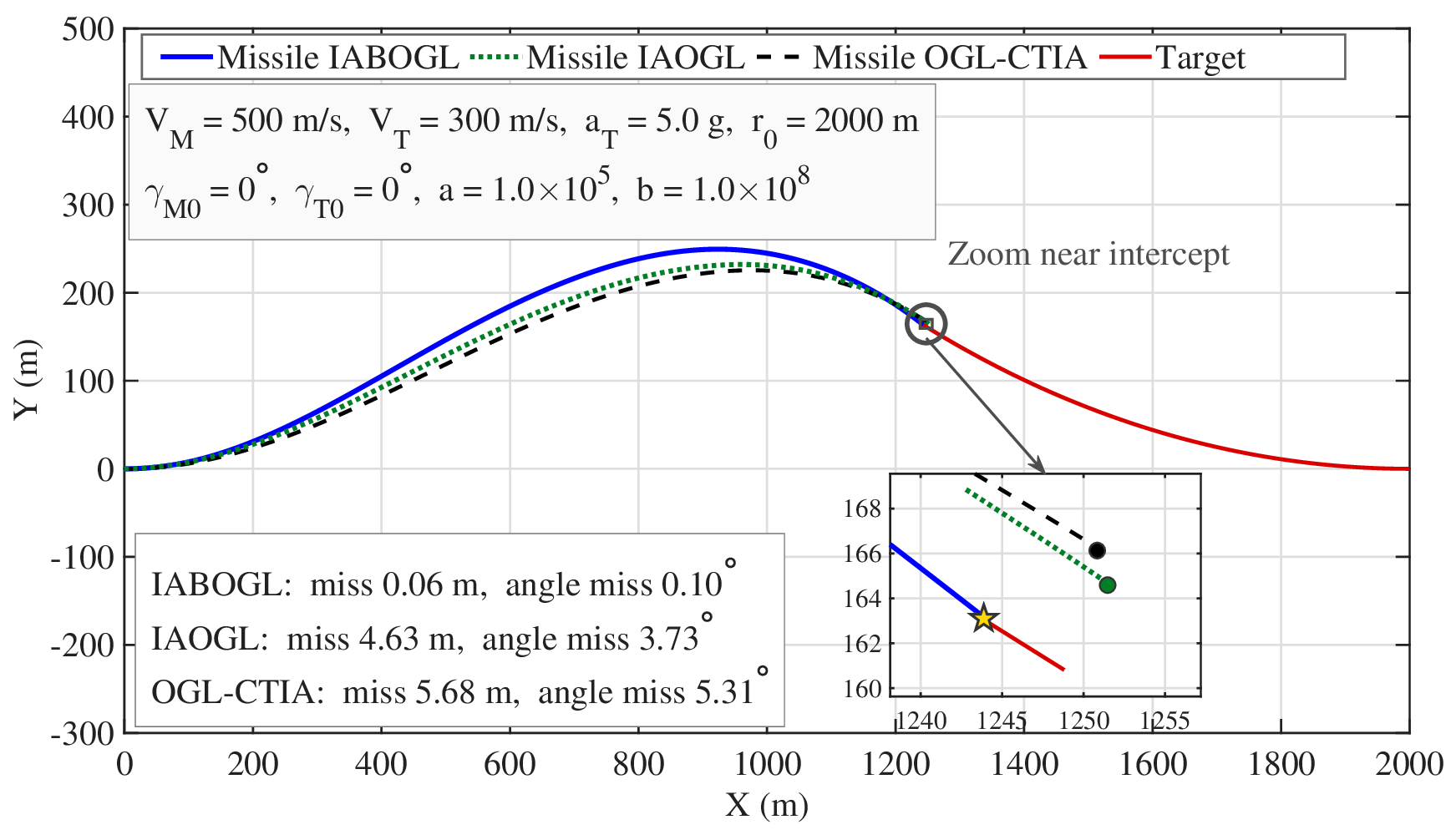}
\caption{\label{fig:case_m6_trajectory} Representative run for \(\chi_c=-6^\circ\) - Trajectories.}
\end{figure}

Figure~\ref{fig:case_m6_trajectory} shows that the IABOGL trajectory intercepts the target with essentially zero miss while preserving the prescribed terminal geometry. In contrast, once command saturation is encountered, both the IAOGL and OGL-CTIA trajectories depart from their unconstrained solutions, accumulating both miss-distance and terminal-angle errors. The zoomed terminal view highlights these deviations during the terminal phase of the engagement.

\begin{figure}[!htbp]
\centering
\includegraphics[width=0.65\textwidth]{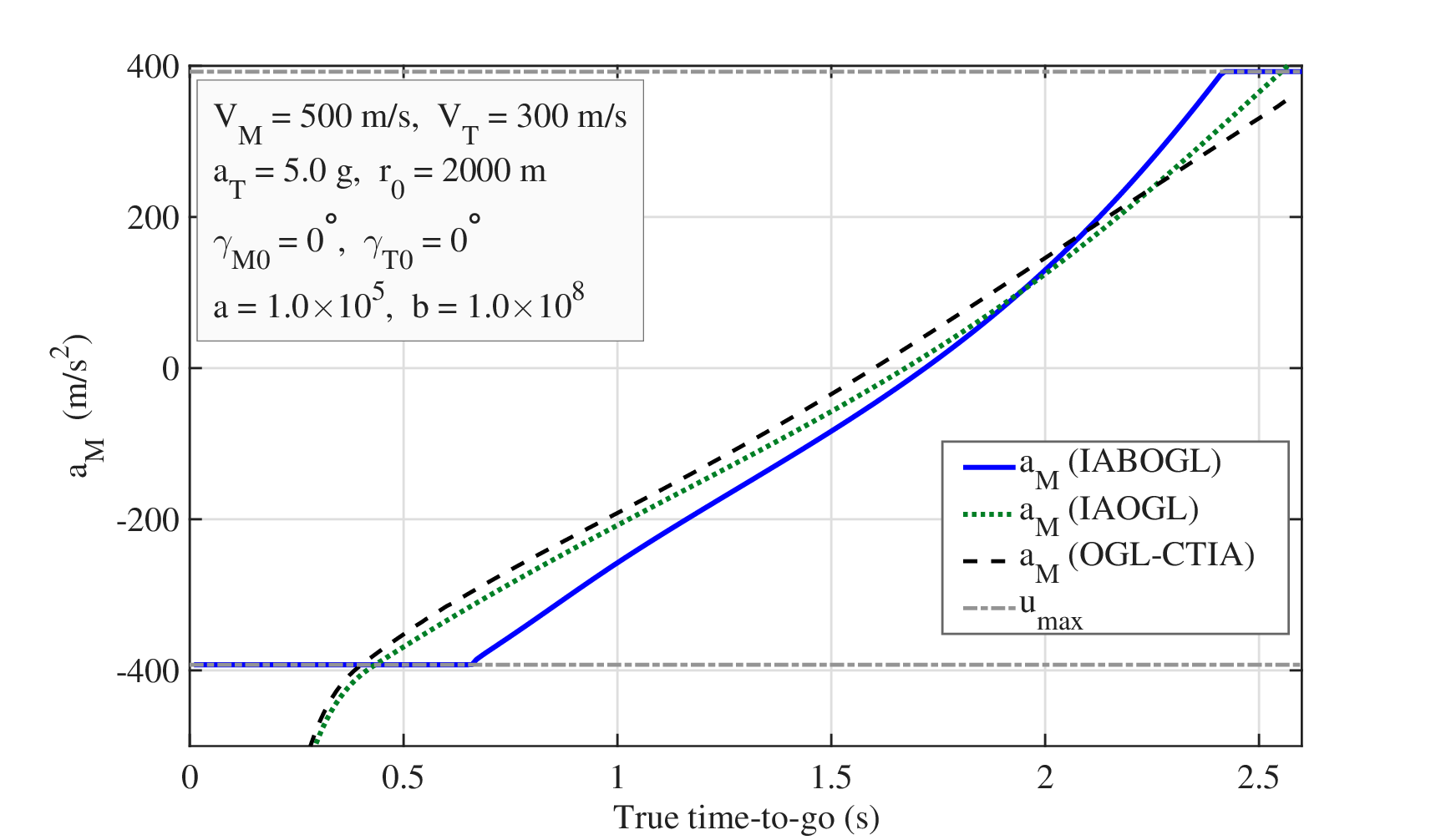}
\caption{\label{fig:case_m6_command} Representative run for \(\chi_c=-6^\circ\) - Acceleration commands.}
\end{figure}

Fig.~\ref{fig:case_m6_command} explains this difference. The IABOGL command saturates at the beginning of the engagement and re-enters the final saturated segment earlier than both the IAOGL and OGL-CTIA. This indicates that the bounded law anticipates the need to use the available acceleration authority at both ends of the horizon, whereas the saturated implementations of the unbounded laws react later because their derivations do not account for the hard acceleration limit.

\begin{figure}[!htbp]
\centering
\includegraphics[width=0.65\textwidth]{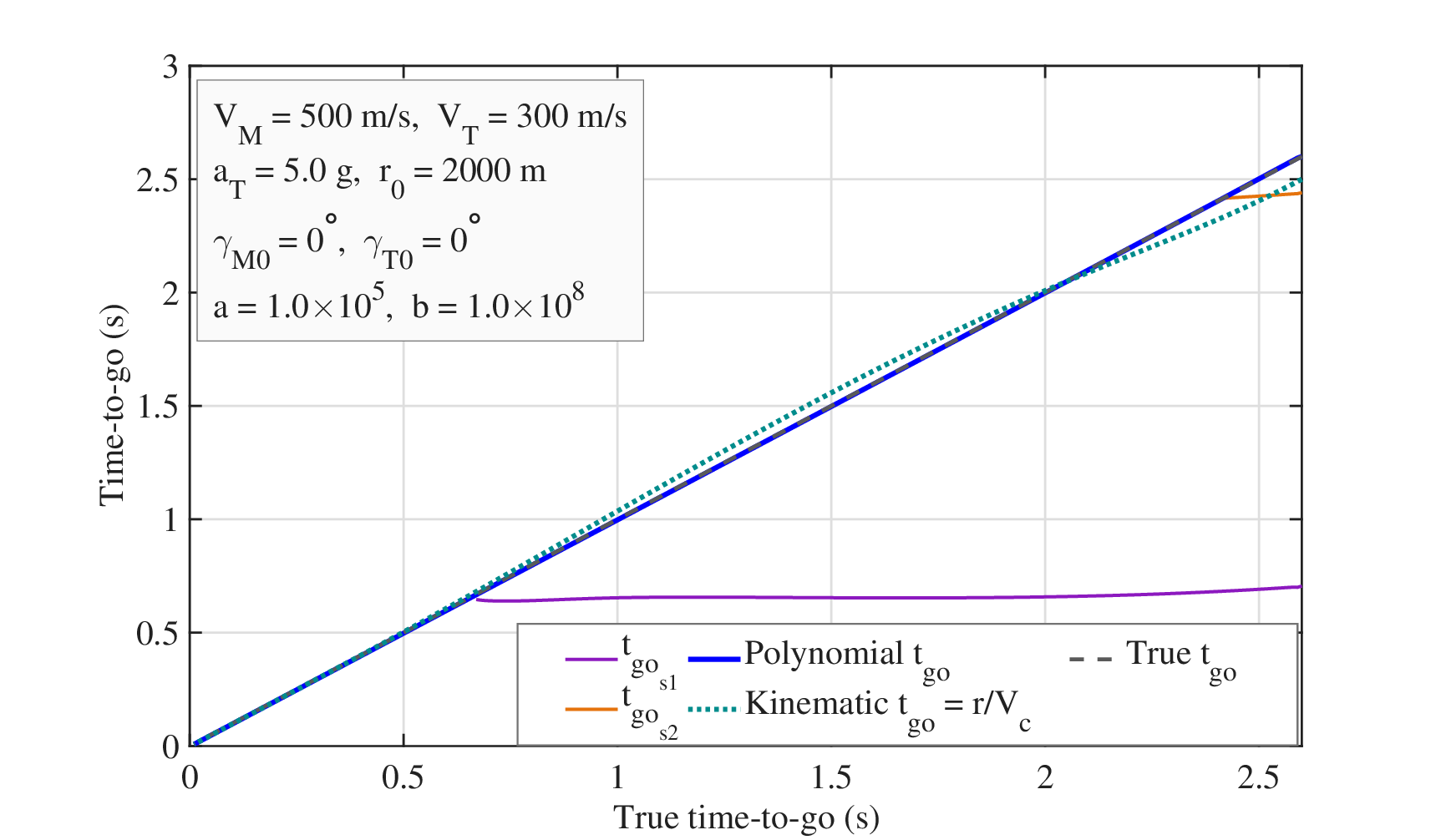}
\caption{\label{fig:case_m6_switch_times} Representative run for \(\chi_c=-6^\circ\) - Switching times history.}
\end{figure}

Figures~\ref{fig:case_m6_command} and~\ref{fig:case_m6_switch_times} show the saturation structure of the bounded solution. The applied command is saturated at the start of the engagement, unsaturated over the middle, and saturated again near intercept. The predicted mode sequence follows \(\mathrm{SUS}\to\mathrm{US}\to\mathrm{S}\), corresponding to an initial two-switch schedule, an intermediate single-switch schedule, and final full saturation near intercept. The nearly constant values of \(t_{go_{s1}}\) and \(t_{go_{s2}}\) over most of the engagement indicate that the local switching structure evolves smoothly from one guidance update to the next and that the guidance law predicts the switching times consistently, despite the nonlinear engagement dynamics and the use of a linear guidance model.

Figure~\ref{fig:case_m6_switch_times} also compares the two time-to-go estimates along the same engagement: the polynomial prediction of \eqref{eq:range_relation_tgo} remains almost identical to the true value, while the kinematic estimate of \eqref{eq:tgo_guess} deviates more noticeably. Substituting the kinematic estimate into the same IABOGL equations, while changing only the time-to-go used in the switching and terminal-value calculations, degrades the terminal accuracy by an order of magnitude; the corresponding results are omitted for brevity.

\begin{figure}[!htbp]
\centering
\includegraphics[width=0.65\textwidth]{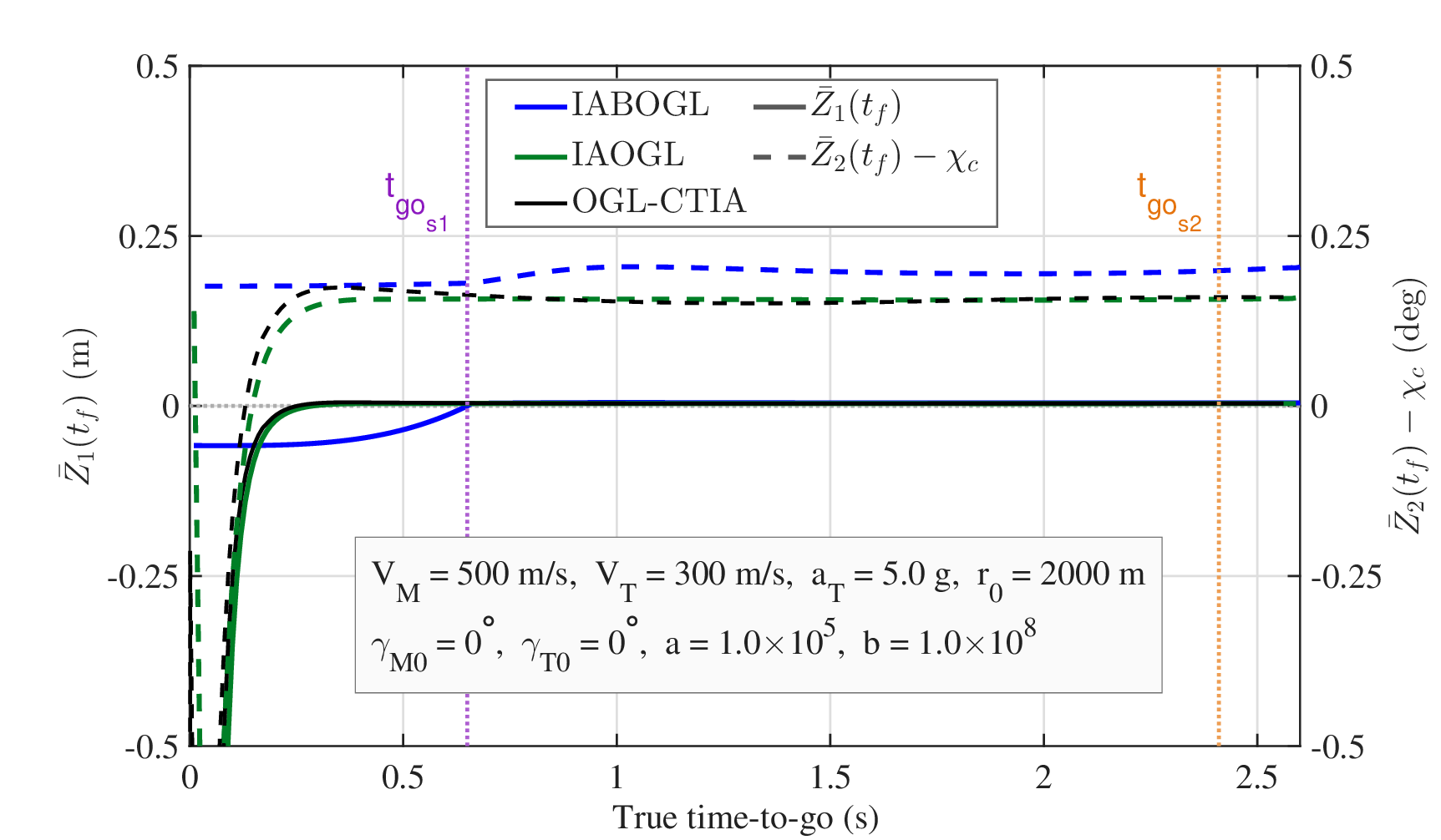}
\caption{\label{fig:case_m6_zbar1_tf_compare} Representative run for \(\chi_c=-6^\circ\) - Predicted terminal values.}
\end{figure}

Figure~\ref{fig:case_m6_zbar1_tf_compare} presents the predicted terminal quantities. Despite saturation, the IABOGL maintains them near their desired values throughout the engagement: \(\bar Z_1(t_f)\) stays close to zero and \(\bar Z_2(t_f)\) remains close to the commanded terminal angle, with the final predicted \(\bar Z_1(t_f)\) corresponding to the achieved miss. The IAOGL and OGL-CTIA predictions begin from the same values and follow each other closely over most of the engagement, but both deviate from their desired values near intercept, where their unconstrained commands exceed the acceleration bound.

\subsection{Parametric Sweep Over the Commanded Terminal Angle}
\label{subsec:parametric_sweep}
\begin{figure}[!htbp]
\centering
\includegraphics[width=0.65\textwidth]{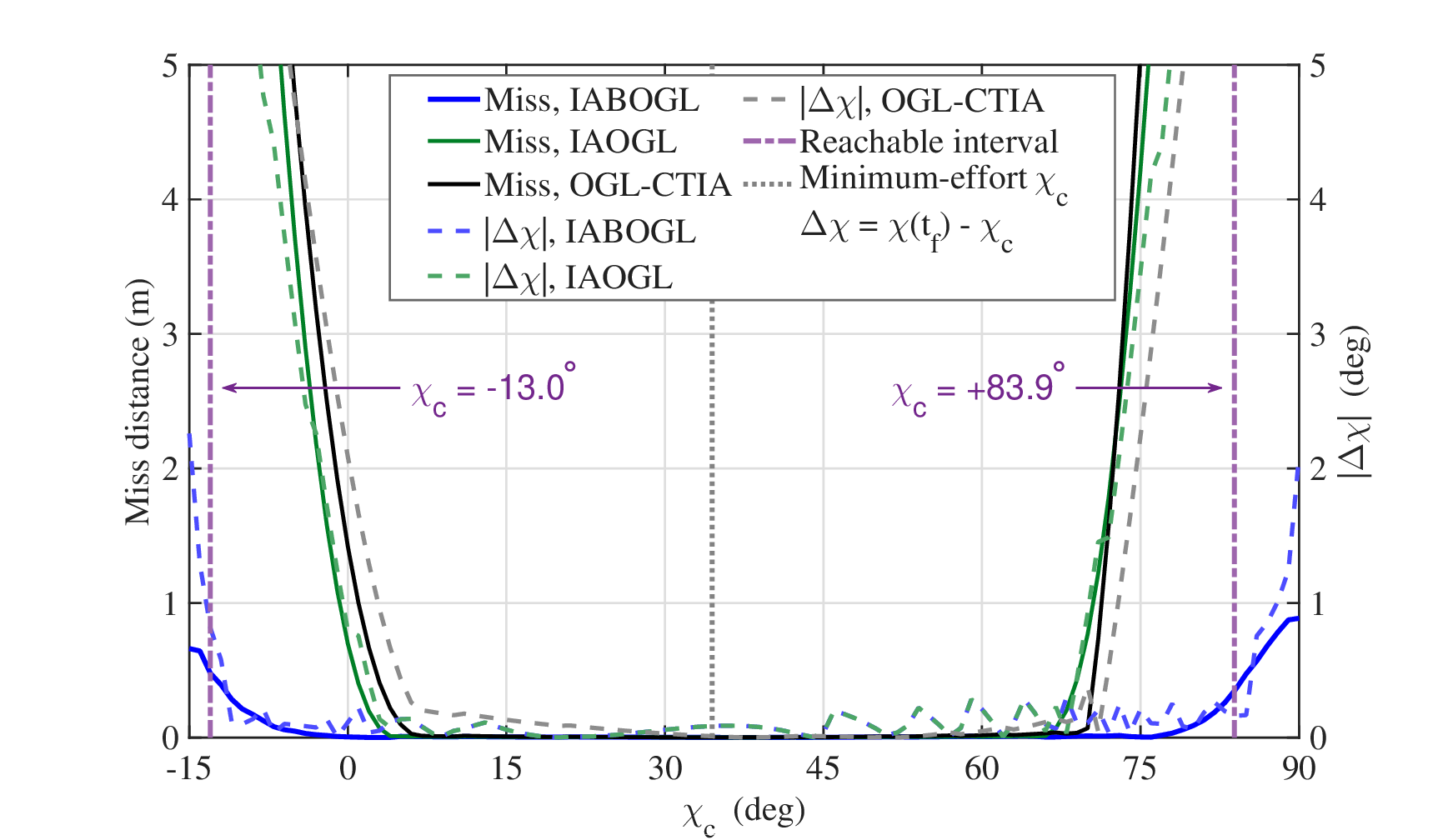}
\caption{\label{fig:sweep_chi_c} Commanded terminal angle sweep - Miss distance and terminal-angle error.}
\end{figure}

To assess performance over a broader range of terminal-angle demands, a sweep over \(\chi_c\) was carried out. Figure~\ref{fig:sweep_chi_c} presents the interval \(\chi_c\in[-15^\circ,+90^\circ]\), over which the transition between successful and unsuccessful cases is visible at both ends. The IABOGL satisfies the miss and angle criteria of \(0.5\) m and \(0.5^\circ\), respectively, from approximately \(\chi_c=-11.5^\circ\) to \(+85^\circ\), whereas the OGL-CTIA satisfies them only from approximately \(+5^\circ\) to \(+70^\circ\). The IAOGL behaves similarly to the OGL-CTIA, satisfying the same criteria only from approximately \(+2^\circ\) to \(+69^\circ\), indicating that the improvement comes from including the bound in the design rather than from the time-varying geometry. The bounded formulation therefore extends the successful region by about \(16^\circ\) toward more demanding negative terminal-angle commands and by about \(15^\circ\) at the positive end. This improvement is concentrated in cases that encounter saturation near both the beginning and end of the engagement, consistent with the representative case discussed earlier. In such cases, the bounded formulation explicitly accounts for the saturated intervals, whereas the saturated OGL-CTIA and IAOGL depart from the command structure assumed in their derivation. Outside the IABOGL successful region, none of the three laws satisfies both criteria under the available acceleration limit. Also shown for reference as a dotted vertical line is \(\chi_c=34.5^\circ\), the terminal angle obtained when the angle weight \(b\) in \eqref{eq:J_in_Z_tv} is set to zero; the smallest control effort and miss distances occur near this value.

Both edges of the IABOGL capture region are consistent with the reachability conditions of Sec.~\ref{subsec:reach_zero_order}. Evaluating these conditions at the initial state of this engagement gives an admissible interval of approximately \([-13.0^\circ,\,+83.9^\circ]\), shown by the vertical lines in Fig.~\ref{fig:sweep_chi_c}. The terminal miss distance and terminal-angle error remain small for milder commands and grow as \(\chi_c\) approaches either limit. At both edges, the observed boundary lies within \(1\)–\(2^\circ\) of the predicted value. The linear reachability conditions therefore predict both boundaries observed in the nonlinear simulation. The miss and angle criteria of \(0.5\) m and \(0.5^\circ\) are arbitrary; relaxing these criteria would expand the observed successful region and make the reachability prediction more conservative.

\section{Conclusions}
\label{sec:conclusions}
A linear-quadratic optimal guidance law for missile interception with a prescribed terminal intercept angle and an explicit hard acceleration-command limit was derived and evaluated in nonlinear simulation. The engagement was formulated as a bounded optimal-control problem and expressed in terms of the zero-effort miss and zero-effort angle, reducing it to a two-state terminal-correction problem whose switching times and terminal values are obtained by solving the per-arc optimality system.

Polynomial approximations of the line-of-sight projection coefficients were used to better represent the nonlinear engagement geometry and determine the time-to-go. The geometry-based time-to-go was found to be an important part of the guidance law, with purely kinematic estimates degrading both the miss distance and terminal-angle performance.

For the zero-order model, the conditions under which the terminal demands can be met were derived in closed form, yielding the minimum and maximum reachable commanded intercept angles. These analytical limits agreed well with those observed in the nonlinear simulation.

The simulations showed that the proposed IABOGL substantially improves both the miss distance and terminal-angle accuracy relative to the unconstrained IAOGL and OGL-CTIA over a broad range of intercept angles when saturation is encountered. The largest improvement occurred in cases involving saturation near both the beginning and end of the engagement, where the bounded law anticipates the upcoming saturation and compensates through earlier maneuvers during the preceding unsaturated interval. Unlike corresponding classical miss-only bounded-guidance formulations for minimum-phase missiles, the proposed intercept-angle-constrained formulation does not reduce to its unconstrained counterpart, since the optimal acceleration command generally does not converge to zero near interception and is typically saturated near the end of the engagement in challenging scenarios.

\appendix
\renewcommand{\thesection}{Appendix \Alph{section}}
\section{Reduced-Order Dynamics and Zero-Effort Variables}
\label{app:reduced_order_ZEV}

This appendix derives the influence functions in Eqs. (\ref{eq:B1B2_closed_form_main}) 
and the closed-form expressions of the zero-effort variables in \eqsref{eq:Z_proj_constaT}.

\subsection{Reduced-Order Dynamics}
Using the time-to-go variable $t_{go}=t_f-t$ and defining $\bfPhi(t_{go})\triangleq\bfPhi(t_f,t)$, \eqref{eq:stm_def_main} becomes
\begin{equation}
\label{eq:Phi_backward_app}
\frac{d\bfPhi(t_{go})}{dt_{go}}=\bfPhi(t_{go})\,\bfA(t_{go}),
\quad
\bfPhi(0)=\bfI
\end{equation}
Let $\bfPhi(t_{go})$ be written as a block matrix, and denote the first and third rows of $\bfPhi(t_{go})$ by $\boldsymbol\phi_1^T$ and $\boldsymbol\phi_3^T$, respectively, with the structure
\begin{equation}
\label{eq:phi13_blocks_app}
\boldsymbol\phi_1^T=
\begin{bmatrix}\phi_{11} & \phi_{12} & \phi_{13} & \boldsymbol\phi_{1T}^T & \boldsymbol\phi_{1M}^T\end{bmatrix}, \quad
\boldsymbol\phi_3^T=
\begin{bmatrix}\phi_{31} & \phi_{32} & \phi_{33} & \boldsymbol\phi_{3T}^T & \boldsymbol\phi_{3M}^T\end{bmatrix}
\end{equation}

Using \eqref{eq:Zdot_reduced_main} together with $\bfb$ from \eqref{eq:LTV_matrices} yields
\begin{subequations}\label{eq:B1B2_via_phi_app}
\begin{equation}
\label{eq:B1_via_phi_app}
\tilde B_1(t_{go})=\boldsymbol\phi_1^T\,\bfb
=-\phi_{12}\,K_M(t_{go})\,d_M
+\phi_{13}\,\frac{d_M}{V_M}
+\boldsymbol\phi_{1M}^T\,\bfb_M
\end{equation}
\begin{equation}
\label{eq:B2_via_phi_app}
\tilde B_2(t_{go})=\boldsymbol\phi_3^T\,\bfb
=-\phi_{32}\,K_M(t_{go})\,d_M
+\phi_{33}\,\frac{d_M}{V_M}
+\boldsymbol\phi_{3M}^T\,\bfb_M
\end{equation}
\end{subequations}

From \eqref{eq:Phi_backward_app} and the structure of $\bfA(t_{go})$ in \eqref{eq:LTV_matrices}, the scalar entries of $\boldsymbol\phi_1^T$ satisfy
\begin{equation}
\label{eq:row1_scalar_app}
\frac{d\phi_{11}}{dt_{go}}=0,\;\phi_{11}(0)=1;\quad
\frac{d\phi_{12}}{dt_{go}}=\phi_{11}, \; \phi_{12}(0)=0;
\quad
\frac{d\phi_{13}}{dt_{go}}=0,\;\phi_{13}(0)=0
\end{equation}
yielding $\phi_{11}=1,\ \phi_{12}=t_{go},\ \phi_{13}=0$. 
The missile and target row blocks $\boldsymbol\phi_{1T}^T, \boldsymbol\phi_{1M}^T$ satisfy
\begin{equation}
\label{eq:phi1TM_ode_app}
\frac{d\boldsymbol\phi_{1T}^T}{dt_{go}}
=K_T(t_{go})\,t_{go}\,\bfc_T^T+\boldsymbol\phi_{1T}^T\bfA_T,
\quad
\boldsymbol\phi_{1T}^T(0)=\mathbf{0}; \quad
\frac{d\boldsymbol\phi_{1M}^T}{dt_{go}}
=-K_M(t_{go})\,t_{go}\,\bfc_M^T+\boldsymbol\phi_{1M}^T\bfA_M,
\quad
\boldsymbol\phi_{1M}^T(0)=\mathbf{0}
\end{equation}
Substituting the polynomial approximations of $K_M(t_{go})$ and $K_T(t_{go})$ from \eqref{eq:KM_KT_polynomials}, applying the Laplace transform, rearranging, using the initial condition, and performing the inverse transformation yields
\begin{equation}
\label{eq:phi1TM_closed_app}
\boldsymbol\phi_{1T}^T(t_{go})
=\sum_{i=0}^{n_T}k_{T,i}(i+1)!\,\mathcal{L}^{-1}_{t_{go}}\!\left[\frac{\bfc_T^T(s\bfI-\bfA_T)^{-1}}{s^{i+2}}\right], \quad
\boldsymbol\phi_{1M}^T(t_{go})
=-\sum_{i=0}^{n_M}k_{M,i}(i+1)!\,\mathcal{L}^{-1}_{t_{go}}\!\left[\frac{\bfc_M^T(s\bfI-\bfA_M)^{-1}}{s^{i+2}}\right]
\end{equation}
Using the same steps for the third row, the scalar entries of $\boldsymbol\phi_3^T$ yield: $\phi_{31}=0,\ \phi_{32}=0,\ \phi_{33}=1$, and the missile and target row blocks $\boldsymbol\phi_{3T}^T, \boldsymbol\phi_{3M}^T$ satisfy
\begin{equation}
\label{eq:phi3T_ode_app}
\frac{d\boldsymbol\phi_{3T}^T}{dt_{go}}=\frac{1}{V_T}\bfc_T^T+\boldsymbol\phi_{3T}^T\bfA_T,
\quad
\boldsymbol\phi_{3T}^T(0)=\mathbf{0}; \quad
\frac{d\boldsymbol\phi_{3M}^T}{dt_{go}}=\frac{1}{V_M}\bfc_M^T+\boldsymbol\phi_{3M}^T\bfA_M,
\quad
\boldsymbol\phi_{3M}^T(0)=\mathbf{0}
\end{equation}
with closed-form solutions
\begin{equation}
\label{eq:phi3T_phi3M_closed_app}
\boldsymbol\phi_{3T}^T(t_{go})
=\frac{1}{V_T}\,\mathcal{L}^{-1}_{t_{go}}\!\left[\frac{\bfc_T^T(s\bfI-\bfA_T)^{-1}}{s}\right], \quad  
\boldsymbol\phi_{3M}^T(t_{go})
=\frac{1}{V_M}\,\mathcal{L}^{-1}_{t_{go}}\!\left[\frac{\bfc_M^T(s\bfI-\bfA_M)^{-1}}{s}\right]
\end{equation}
Substituting the values of $\boldsymbol\phi_1^T$, $\boldsymbol\phi_3^T$ into \eqref{eq:B1B2_via_phi_app}, and using $G_M(s)=\bfc_M^T(s\bfI-\bfA_M)^{-1}\bfb_M+d_M$ from \eqref{eq:GM_def}, yields \eqref{eq:B1B2_closed_form_main}.

\subsection{Zero-Effort Variables}
From \eqref{eq:Z_def_main}, $Z_i=Z_i^H+Z_i^T , \ i\in\{1,2\}$, where
\begin{equation}
\label{eq:ZH_app}
\begin{split}
Z_1^H=\boldsymbol\phi_1^T(t_{go})\,\bfx(t),
\quad
Z_1^T=\int_t^{t_f}\boldsymbol\phi_1^T(t_f-\tau)\,\bfc(\tau)\,w(\tau)\,d\tau \\
Z_2^H=\boldsymbol\phi_3^T(t_{go})\,\bfx(t),
\quad
Z_2^T=\int_t^{t_f}\boldsymbol\phi_3^T(t_f-\tau)\,\bfc(\tau)\,w(\tau)\,d\tau
\end{split}
\end{equation}
The homogeneous components follow directly from $\boldsymbol\phi_1^T$ and $\boldsymbol\phi_3^T$
\begin{equation}
\label{eq:Z1H_app}
Z_1^H
=x_1+t_{go}\,x_2
-\sum_{i=0}^{n_M}k_{M,i}(i+1)!\,\mathcal{L}^{-1}_{t_{go}}\!\left[\frac{\bfc_M^T(s\bfI-\bfA_M)^{-1}}{s^{i+2}}\right]\bfx_M
+\sum_{i=0}^{n_T}k_{T,i}(i+1)!\,\mathcal{L}^{-1}_{t_{go}}\!\left[\frac{\bfc_T^T(s\bfI-\bfA_T)^{-1}}{s^{i+2}}\right]\bfx_T
\end{equation}
\begin{equation}
\label{eq:Z2H_app}
Z_2^H
=x_3
+\frac{1}{V_M}\,\mathcal{L}^{-1}_{t_{go}}\!\left[\frac{\bfc_M^T(s\bfI-\bfA_M)^{-1}}{s}\right]\bfx_M
+\frac{1}{V_T}\,\mathcal{L}^{-1}_{t_{go}}\!\left[\frac{\bfc_T^T(s\bfI-\bfA_T)^{-1}}{s}\right]\bfx_T
\end{equation}
The target-driven components, by the same procedure, using $G_T(s)=\bfc_T^T(s\bfI-\bfA_T)^{-1}\bfb_T+d_T$, and noting that $Z_i^T$ are convolutions yield
\begin{equation}
\label{eq:Z1TZ2T_app}
Z_1^T
=\sum_{i=0}^{n_T}k_{T,i}(i+1)!\,\mathcal{L}^{-1}_{t_{go}}\!\left[\frac{G_T(s)\,\tilde w(s)}{s^{i+2}}\right], \quad
Z_2^T
=\frac{1}{V_T}\,\mathcal{L}^{-1}_{t_{go}}\!\left[\frac{G_T(s)\,\tilde w(s)}{s}\right]
\end{equation}
Combining \eqref{eq:Z1H_app}--\eqref{eq:Z1TZ2T_app}, rearranging, and using
\(\tilde a_T(s)=G_T(s)\tilde w(s)+\bfc_T^T(s\bfI-\bfA_T)^{-1}\bfx_T\),
which accounts for the initial conditions, yields
\begin{subequations}\label{eq:Z1Z2_general_app}
\begin{equation}
\label{eq:Z1_general_app}
Z_1
=x_1+t_{go}\,x_2
-\sum_{i=0}^{n_M}k_{M,i}(i+1)!\,\mathcal{L}^{-1}_{t_{go}}\!\left[\frac{\bfc_M^T(s\bfI-\bfA_M)^{-1}}{s^{i+2}}\right]\bfx_M
+\sum_{i=0}^{n_T}k_{T,i}(i+1)!\,\mathcal{L}^{-1}_{t_{go}}\!\left[\frac{\tilde a_T(s)}{s^{i+2}}\right]
\end{equation}
\begin{equation}
\label{eq:Z2_general_app}
Z_2
=x_3
+\frac{1}{V_M}\,\mathcal{L}^{-1}_{t_{go}}\!\left[\frac{\bfc_M^T(s\bfI-\bfA_M)^{-1}}{s}\right]\bfx_M
+\frac{1}{V_T}\,\mathcal{L}^{-1}_{t_{go}}\!\left[\frac{\tilde a_T(s)}{s}\right]
\end{equation}
\end{subequations}
For constant future target maneuver $a_T$, $\tilde a_T(s)=a_T/s$, and $\mathcal{L}^{-1}_{t_{go}}[a_T/s^{i+3}]=a_T t_{go}^{i+2}/(i+2)!$. The target term in \eqref{eq:Z1_general_app} simplifies as $k_{T,i}(i+1)!\cdot a_T t_{go}^{i+2}/(i+2)!=k_{T,i}\,a_T\,t_{go}^{i+2}/(i+2)$, and the target term in \eqref{eq:Z2_general_app} becomes $a_T t_{go}/V_T$, giving
\begin{subequations}\label{eq:Z1Z2_constaT_app}
\begin{equation}
\label{eq:Z1_constaT_app}
Z_1
=x_1+t_{go}\,x_2
-\sum_{i=0}^{n_M}k_{M,i}(i+1)!\,\mathcal{L}^{-1}_{t_{go}}\!\left[\frac{\bfc_M^T(s\bfI-\bfA_M)^{-1}}{s^{i+2}}\right]\bfx_M
+\sum_{i=0}^{n_T}\frac{k_{T,i}\,a_T}{i+2}\,t_{go}^{\,i+2}
\end{equation}
\begin{equation}
\label{eq:Z2_constaT_app}
Z_2
=x_3
+\frac{1}{V_M}\,\mathcal{L}^{-1}_{t_{go}}\!\left[\frac{\bfc_M^T(s\bfI-\bfA_M)^{-1}}{s}\right]\bfx_M
+\frac{a_T\,t_{go}}{V_T}
\end{equation}
\end{subequations}
which concludes the derivation of \eqsref{eq:Z_proj_constaT}.

\section{Inner-Solver Jacobian}
\label{app:jacobian}

This appendix gives the closed-form Jacobian of the inner-solver system
\eqref{eq:Fsystem} used in Sec.~\ref{subsec:inner_solver}, and the row that replaces the
last switching condition at a mode boundary.

\subsection{Terminal States and Switching-Condition Rows}
\label{app:jacobian_rows}

Differentiating the terminal state equations \(E_1\) and \(E_2\) with respect to \(\bar Z_1(t_f)\), \(\bar Z_2(t_f)\) yields
\begin{equation}
\label{eq:leibniz_col}
\frac{\partial E_1}{\partial \bar Z_1(t_f)}=1+a\bar{\mathcal I}_{11}, \quad
\frac{\partial E_1}{\partial \bar Z_2(t_f)}= b\bar{\mathcal I}_{12}, \quad
\frac{\partial E_2}{\partial \bar Z_1(t_f)}=a\bar{\mathcal I}_{12}, \quad
\frac{\partial E_2}{\partial \bar Z_2(t_f)}= 1+b\bar{\mathcal I}_{22}
\end{equation}
In \(E_1\) and \(E_2\) the switching times appear only as the limits of the arc integrals
\(\bar{\mathcal I}_{ij}\) and the saturated contributions \(\mathcal S_1,\mathcal S_2\). With
\(\mathcal Q_j(\xi)=\int_0^\xi\tilde B_j\,d\xi'\) and
\(\mathcal I_{ij}(\xi)=\int_0^\xi\tilde B_i\tilde B_j\,d\xi'\), the derivatives are
\(\mathcal Q_j'(c)=\tilde B_j(c)\) and \(\mathcal I_{ij}'(c)=\tilde B_i(c)\tilde B_j(c)\).
Differentiating the terminal state equations \(E_1\) and \(E_2\) with respect to a switching time then gives
\begin{equation}
\label{eq:leibniz_col_simple}
\frac{\partial E_i}{\partial t_{go_{sk}}}=\bar{s}_k\,\tilde B_i(t_{go_{sk}})\bigl[\bar u_0(t_{go_{sk}})-S_k u_{\max}\bigr],\quad i\in\{1,2\}
\end{equation}
where \(\bar{s}_k=+1\) if the arc just before \(t_{go_{sk}}\), toward intercept, is saturated and
\(\bar{s}_k=-1\) if it is unsaturated. These vanish at a solution, where
\(\bar u_0(t_{go_{sk}})=S_k u_{\max}\). Note that \(\bar{s}_k\) indicates the type of the preceding arc, whereas \(S_k\) denotes the sign of the saturated command at the switch.

The switching rows differentiate \(E_{2+k}=\bar u_0(t_{go_{sk}})-S_k u_{\max}\), giving
\begin{equation}
\label{eq:switch_row_partials}
\frac{\partial E_{2+k}}{\partial\bar Z_1(t_f)}=-a\,\tilde B_1(t_{go_{sk}}),\quad
\frac{\partial E_{2+k}}{\partial\bar Z_2(t_f)}=-b\,\tilde B_2(t_{go_{sk}}),\quad
\frac{\partial E_{2+k}}{\partial t_{go_{sk}}}=\bar u_0'(t_{go_{sk}})
\end{equation}
with
\begin{equation}
\label{eq:u0_prime}
\bar u_0'(\xi)=-a\,\bar Z_1(t_f)\,\tilde B_1'(\xi)-b\bigl[\bar Z_2(t_f)-\chi_c\bigr]\tilde B_2'(\xi),
\quad
\tilde B_i'(\xi)= \frac{\partial \tilde B_i(\xi)}{\partial \xi}
\end{equation}
and the remaining switching entries zero. The full \((2+m)\times(2+m)\) Jacobian is
\begin{equation}
\label{eq:jacobian_full}
\mathbf J=\begin{bmatrix}
1+a\bar{\mathcal I}_{11} & b\bar{\mathcal I}_{12} & \dfrac{\partial E_1}{\partial t_{go_{s1}}} & \cdots & \dfrac{\partial E_1}{\partial t_{go_{sm}}}\\[6pt]
a\bar{\mathcal I}_{12} & 1+b\bar{\mathcal I}_{22} & \dfrac{\partial E_2}{\partial t_{go_{s1}}} & \cdots & \dfrac{\partial E_2}{\partial t_{go_{sm}}}\\[6pt]
-a\tilde B_1(t_{go_{s1}}) & -b\tilde B_2(t_{go_{s1}}) & \bar u_0'(t_{go_{s1}}) & &\\
\vdots & \vdots & & \ddots &\\
-a\tilde B_1(t_{go_{sm}}) & -b\tilde B_2(t_{go_{sm}}) & & & \bar u_0'(t_{go_{sm}})
\end{bmatrix}
\end{equation}
At a solution obtained with the ordinary switching conditions, the terminal-row switching
columns \eqref{eq:leibniz_col_simple} vanish, so \(\mathbf J\) is block lower-triangular. 
This block structure does not hold for a candidate solved with the mode-boundary row of \ref{app:fb_row}, where the command at \(t_{go_{sm}}\) is no longer on the bound.

In the implementation, the switching rows of \(\mathbf F\) and the corresponding rows of
\(\mathbf J\) are scaled by \(1/u_{\max}\), so that the residual and the Jacobian row carry
the same factor and it cancels in the Newton step. The scaling leaves the step unchanged and
keeps the components of \(\mathbf F\) comparable in magnitude.

\subsection{Mode-Boundary Row}
\label{app:fb_row}
As stated in Sec.~\ref{subsubsec:mode_transitions}, the last switching condition \(E_{2+m}\) is replaced by the Fischer--Burmeister condition \(\varphi(p,q)=0\) \eqref{eq:phi} for every candidate with \(m\ge1\), with \(p\) and \(q\) given by \eqref{eq:pq_def}. The derivatives below are written for a candidate whose arc at \(\xi=t_{go}\) is saturated, in which case arc \(m\) is unsaturated and \(q\) is the first branch of \eqref{eq:pq_def}; the two branches differ only in sign, so in the other case every derivative of \(q\) changes sign, while \(\partial p/\partial t_{go_{sm}}\) is unchanged. The partial derivatives of \(\varphi\) are
\begin{equation}
\label{eq:phi_pq_derivs}
\frac{\partial\varphi}{\partial p}=1-\frac{p}{\sqrt{p^2+q^2}},\quad
\frac{\partial\varphi}{\partial q}=1-\frac{q}{\sqrt{p^2+q^2}}
\end{equation}
and, from \eqref{eq:pq_def} together with \(\bar u_0\) in \eqref{eq:bounded_command_general}
\begin{subequations}
\label{eq:pq_chain}
\begin{align}
\frac{\partial p}{\partial t_{go_{sm}}}&=-\frac{1}{t_{go}},\quad
\frac{\partial q}{\partial t_{go_{sm}}}=-\frac{S_m}{u_{\max}}\,\bar u_0'(t_{go_{sm}})\\
\frac{\partial q}{\partial\bar Z_1(t_f)}&=\frac{S_m\,a}{u_{\max}}\,\tilde B_1(t_{go_{sm}}),\quad
\frac{\partial q}{\partial\bar Z_2(t_f)}=\frac{S_m\,b}{u_{\max}}\,\tilde B_2(t_{go_{sm}})
\end{align}
\end{subequations}
with \(p\) independent of the terminal values. The chain rule gives the replaced row
\begin{subequations}
\label{eq:fb_row}
\begin{align}
\frac{\partial\varphi}{\partial\bar Z_1(t_f)}
&=\frac{\partial\varphi}{\partial q}\,\frac{S_m\,a}{u_{\max}}\,\tilde B_1(t_{go_{sm}})\\
\frac{\partial\varphi}{\partial\bar Z_2(t_f)}
&=\frac{\partial\varphi}{\partial q}\,\frac{S_m\,b}{u_{\max}}\,\tilde B_2(t_{go_{sm}})\\
\frac{\partial\varphi}{\partial t_{go_{sm}}}
&=-\frac{\partial\varphi}{\partial p}\,\frac{1}{t_{go}}
-\frac{\partial\varphi}{\partial q}\,\frac{S_m}{u_{\max}}\,\bar u_0'(t_{go_{sm}})
\end{align}
\end{subequations}
The remaining entries of the row are zero, since \(\varphi\) does not depend on the other
switching times. The row is dimensionless and is not rescaled.

\section*{Acknowledgment}
This work was supported by the PMRI – Peter Munk Research Institute - Technion. 

The authors acknowledge using ChatGPT and Grammarly for editorial assistance during manuscript preparation, including grammar correction and language refinement.

\bibliography{references_2}

\end{document}

%% file: command.tex
\newcommand{\bfb}{\boldsymbol{b}}

\newcommand{\bfPhi}{\mathbf{\Phi}}

\newcommand{\bfA}{\mathbf{A}}

\newcommand{\bfx}{\mathbf{x}}

\newcommand{\bfI}{\mathbf{I}}

\newcommand{\bfZ}{\mathbf{Z}}
\newcommand{\bfB}{\mathbf{B}}

\newcommand{\bfc}{\mathbf{c}}

\newcommand{\bfE}{\mathbf{E}}

\renewcommand{\eqref}[1]{Eq.~(\ref{#1})}
\newcommand{\eqsref}[1]{Eqs.~(\ref{#1})}

